\documentclass[aps,prb,twocolumn,showpacs]{revtex4}
\usepackage{amsmath}
\usepackage{amsthm}
\usepackage{amssymb}
\usepackage{graphicx}
\usepackage{tabularx}
\usepackage[squaren]{SIunits}

\begin{document}
\title{Local moment formation and Kondo screening in impurity trimers}
\date{\today}

\author{Andrew K. Mitchell}
\author{Thomas F. Jarrold}
\author{Martin R. Galpin}
\author{David E. Logan}
\affiliation{Department of Chemistry, Physical and Theoretical Chemistry,
Oxford University, South Parks Road, Oxford OX1 3QZ, United Kingdom}


\begin{abstract}
We study theoretically a triangular cluster of three magnetic impurities, hybridizing locally with conduction electrons of a metallic host. Such a cluster is the simplest to exhibit frustration -- an important generic feature of many complex molecular systems in which different interactions compete. Here, low-energy doublet states of the trimer are favored by effective exchange interactions produced by strong electronic repulsion in localized impurity orbitals. Parity symmetry protects a level-crossing of such 
states on tuning microscopic parameters, while an avoided crossing arises in the general distorted case. On coupling to a metallic host, the behavior is shown to be immensely rich, since collective quantum many-body effects now also compete. In particular, impurity degrees of freedom are totally screened at low temperatures in a Kondo-screened Fermi liquid phase, while degenerate ground states persist in a local moment phase. Local frustration
drives the quantum phase transition between the two, which may be first order or of Kosterlitz-Thouless type, depending on symmetries. Unusual mechanisms for local moment formation and Kondo screening are found, due to the 
orbital structure of the impurity trimer. 
Our results are of relevance for triple quantum dot devices. The problem is studied by a combination of analytical arguments and the numerical renormalization group.
\end{abstract}

\maketitle


\section{Introduction and motivation}\label{sec:intro}

The interplay of orbital and spin degrees of freedom with electronic interactions can produce a diverse range of chemical and physical behavior. At the few-electron level of a single molecule, 
understanding the resulting complexity is a traditional 
problem in theoretical chemistry --- and a challenge 
because strong correlations preclude an independent particle picture.
At the many-electron level of clean bulk metals, by contrast, interactions are often rather unimportant, and the system adequately described by an essentially independent particle description.

Bridging between these limits is the fascinating class of `quantum
impurity problems', \cite{hewson} in which
an interacting and, in effect, small molecular subsystem is coupled locally to the continuum of conduction
electrons in a metal. 
The most basic example, a classic paradigm in condensed
matter science,\cite{hewson,kondo,siam} 
is a single magnetic impurity  
embedded in a metallic host. Local Coulomb repulsion 
favors single-occupancy 
of the active impurity orbital, with a local moment thus forming 
at high temperatures due to the free spin degree of freedom. But the Kondo 
effect plays a key role at low temperatures/energies: the impurity moment 
is screened dynamically by conduction electrons, which together form a 
many-body spin-singlet state (the `Kondo singlet'). The screening process itself can be 
understood in terms of a renormalization group flow, 
corresponding to the crossover from the local moment fixed point applicable at high temperatures, to the strong coupling fixed point describing the Kondo singlet ground state,\cite{hewson,wilson,kww,akm:rginrs} and with physical properties exhibiting universality in terms of the crossover temperature scale, $T_K$. 
Interestingly, there is a close connection between spin Kondo physics and dissipative tunneling relevant to electron transfer processes in chemistry, as emphasised many years ago by Jongeward and Wolynes in Ref.~\cite{wolynes}.

The Kondo effect associated with screening of a single magnetic impurity has been observed experimentally in many metals.\cite{hewson} In semi-metallic systems by contrast, where the host density of states vanishes at the Fermi level, it is well known that Kondo physics can be suppressed,
and that such systems also support degenerate local moment ground states.\cite{fradkin,sg_ingersent,sg_lma,sg_bulla,sg_lma2} Local quantum phase transitions between Kondo screened phases and local moment phases have thus attracted much interest and have been studied in detail; being sought for example in softgapped systems such as graphene,\cite{rev:graphene,akm:graph} d-wave superconductors\cite{dwavesc} and surfaces of 3d topological insulators.\cite{zitko_ti,akm:qpiti}

The situation is unsurprisingly richer when several impurities are present.  Already in the case of two impurities, the resulting behavior can be markedly different.\cite{2imp_krish,jones,CFT2IKM,dd_rev,Chang_rev,simondimer,akm:ccdqd,DEL_2iam,akm:exactNFL,akm:finiteT,akm:2ckin2ik,del_cntdqd} If the impurities are spatially well separated, each is essentially screened independently by the Kondo effect. But an effective exchange interaction (`RKKY') can arise and dominate when two impurities are brought closer, with a resulting crossover to \emph{local} interimpurity spin-singlet formation\cite{2imp_krish,jones} (indeed for strictly independent screening channels, the crossover is 
sharpened to a true transition\cite{CFT2IKM,akm:exactNFL,akm:finiteT,akm:2ckin2ik}).   

Multi-impurity systems, and even small molecules deposited on metallic surfaces, 
exhibit more complex phenomena due to competition between local interactions and the Kondo effect.\cite{cr3:expt,cr3:theory,Cotrimer,lazarovitsadatom,akm:tqd1ch,akm:tqd2ch,wangtqd,wangtqd2,zitkoTQD2ch,hewsontqd1,hewsontqd2,tqdexpt} In this paper we consider a cluster of three impurities, which exhibits the frustrating effects of competing ground states. 
Frustration and degenerate ground states at level crossings are of course an important general feature of many complex molecular systems familiar 
in chemistry. Here we examine the interplay between frustration and the Kondo effect in one of its simplest realizations, to obtain a detailed understanding of the underlying physical behavior.

Real three-impurity systems, such as Cr clusters on clean Au(111) surfaces or Co clusters on Cu(100), have been studied both experimentally\cite{cr3:expt,Cotrimer} and theoretically.\cite{cr3:theory,lazarovitsadatom,cr3:uzdin,cr3:Gotsis,cr3:frust} But since the impurities couple to different surface sites of the host, a realistic theoretical model maps irreducibly to a three-channel problem,\cite{cr3:theory} whose properties depend quite sensitively on details of the particular experimental realization. More general aspects of the three impurity problem can however be accessed in triple quantum dot devices,\cite{schroer,vidan3dot,rogge3dot,Gaudreau,tqdexpt} where the metallic leads to which the dots are tunnel-coupled provide the conduction electrons.
The Kondo effect has been observed in single semiconductor quantum dots coupled to metallic leads,\cite{qd:expt} which are often referred to as `artificial atoms'.\cite{artatom} By the same token,
coupled dot devices 
behave as highly tunable `artificial molecules'.\cite{artmol,blick,vidan3dot} 
The primary experimental probe in all cases is of course the conductance, obtained by driving a steady current through the lead-coupled quantum dot system.

\begin{figure}
\begin{center}
\includegraphics[height=40mm]{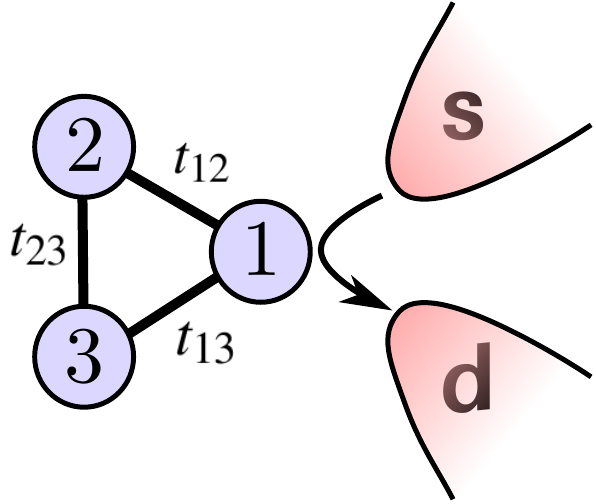}
\caption{\label{fig:tqd} Schematic illustration of the TQD setup. Transport measurements between source (s) and drain (d) leads indicated by the arrow.}
\end{center}
\end{figure}

The system studied here 
is illustrated in Fig.~\ref{fig:tqd}, and consists of a triangular cluster of three quantum dots (impurities), tunnel-coupled apically to a single channel of conduction electrons (and with the metallic lead  split into source and drain to allow measurement of conductance). We find both a Kondo screened phase and a free local moment phase, depending on model parameters (see Fig.~\ref{fig:pd}), which themselves could be tuned in a real device by application of gate voltages.\cite{schroer,vidan3dot,rogge3dot,Gaudreau,tqdexpt} The Kondo phase is characterized by a strong coupling Fermi liquid ground state, in which all dot degrees of freedom are screened. Zero-bias conductance through the device in this case is maximally enhanced at low temperatures due to the Kondo effect. Three distinct screening mechanisms are uncovered, depending on the relative strength of interdot tunnel-couplings, which have characteristic universal signatures in physical quantities. By contrast, the Kondo effect is totally suppressed in the local moment phase, which has a doubly-degenerate ground state and extremely low conductance.

The quantum phase transition separating the Kondo and local moment phases is studied in some detail. In the parity-symmetric case, the transition is a 
parity-protected level crossing.\cite{akm:tqd1ch} 
In the more general distorted case by contrast, the transition occurs at the critical end point of a \emph{line} of Kondo screened states. An effective model is derived to describe this Kosterlitz-Thouless transition,\cite{kt} which captures the characteristic vanishing of the low-energy scale, $T_K$, as the transition is approached from the Kondo phase. Analytic arguments are supplemented and confirmed by exact numerics obtained using the numerical renormalization group.\cite{wilson}


\section{Models and methods}\label{sec:model}

We consider the triple quantum dot (TQD) device illustrated in Fig.~\ref{fig:tqd}. It consists of three equivalent and locally correlated single-level (`Andersonian') sites,
 with level energy $\epsilon$ and on-site Coulomb repulsion $U$.
 Dots $i$ and $j$ are tunnel-coupled by a matrix element $t_{ij}$ to form a triangular arrangement. Dot 1 is also coupled to source and drain leads (and we consider the zero-bias case where the system is in equilibrium). The Hamiltonian is decomposed as $H=H_{\text{leads}}+H_{\text{TQD}}+H_{\text{hyb}}$ where, in standard notation, 
\begin{subequations}\label{eq:H}
\begin{align}
\label{eq:Hleads}
H_{\text{leads}} =& \sum_{\alpha=s,d}\sum_{\sigma=\uparrow,\downarrow} \sum_{\textbf{k}} \epsilon_{\textbf{k}}^{\phantom\dagger} c_{\alpha\textbf{k}\sigma}^{\dagger} c_{\alpha\textbf{k}\sigma}^{\phantom{\dagger}} \;\\
\label{eq:Htqd}
H_{\text{TQD}} =& \sum_{i=1,2,3} \left [ \epsilon ( \hat{n}_{i\uparrow}+\hat{n}_{i\downarrow}) + U \hat{n}_{i\uparrow}\hat{n}_{i\downarrow} \right ] \\
&+ \sum_{i<j, \sigma} \left [ t_{ij}^{\phantom{\dagger}} d_{i\sigma}^{\dagger} d_{j\sigma}^{\phantom{\dagger}} + \text{H.c.}  \right ] \;\\
\label{eq:Hhyb}
H_{\text{hyb}} =& \sum_{\alpha,\textbf{k},\sigma} \left [ V_{\alpha\textbf{k}}^{\phantom{\dagger}} d_{1\sigma}^{\dagger} c_{\alpha \textbf{k} \sigma}^{\phantom{\dagger}} + \text{H.c.} \right ] \;
\end{align}
\end{subequations}
with $\hat{n}_{i\sigma}=d_{i\sigma}^{\dagger} d_{i\sigma}^{\phantom{\dagger}}$ the number operator for spin-$\sigma=\uparrow / \downarrow$ electrons on dot site $i=1,2,3$. 
$H_{\text{TQD}}$ describes the isolated TQD, by itself a 
small quasi-molecular entity that can accommodate up to 6 electrons. In the absence of electron interactions ($U=0$) it is in fact merely a 3-site H\"uckel ring, although interactions are 
essential, since $U$ is inversely proportional to the dot capacitance (and no quantum dot has infinite capacitance!). $H_{\text{leads}}$ by contrast describes the leads, an essentially non-interacting
but macroscopic metal.  
The hybridization term $H_{\text{hyb}}$, in connecting the two subsystems via tunnel-coupling between dot 1 and the leads, ensures 
that the lead-coupled TQD system is an 
interacting, many-body problem containing macroscopic numbers of electrons. That is essential to the basic physics (despite some naive attempts 
to avoid it in the chemistry literature).

For equivalent leads $\alpha=s,d$,
we define the conventional `local' orbital to which dot 1 couples as,
\begin{equation}\label{eq:f0}
f_{0\sigma} = \frac{1}{V}\sum_{\alpha,\textbf{k}} V_{\alpha\textbf{k}} c_{\alpha \textbf{k} \sigma} \qquad ;\qquad V^2 = \sum_{\alpha ,\textbf{k}} V_{\alpha\textbf{k}}^2\;.
\end{equation}
With this, the full TQD problem can then be mapped to an effective 1d problem by tridiagonalizing the conduction electrons,\cite{hewson,wilson} starting from the $f_{0\sigma}$ `zero' orbital. One thereby obtains,
\begin{subequations}
\begin{align}
\label{eq:Hleads2}
H_{\text{leads}} =& \sum_{\sigma}\sum_{n=0}^{\infty} \left [ e_{n} f_{n\sigma}^{\dagger} f_{n\sigma}^{\phantom{\dagger}} + h_n\left ( f_{n\sigma}^{\dagger} f_{(n+1)\sigma} + \text{H.c.} \right) \right ]\\
\label{eq:Hhyb2}
H_{\text{hyb}} =& V \sum_{\sigma} \left [ d_{1\sigma}^{\dagger} f_{0\sigma}^{\phantom{\dagger}} + \text{H.c.} \right ] \;,
\end{align}
\end{subequations}
with all diagonal one-electron energies $e_{n}=0$ 
if the local conduction electron density of states (DOS), $\rho_{\sigma}^{(0)}(\omega)$, is particle-hole symmetric;
with $\rho_{\sigma}^{(0)}(\omega)$ given by
\begin{equation}\label{eq:dos}
\rho_{\sigma}^{(0)}(\omega) =
\sum _{\alpha,\textbf{k}} \left (\frac{V_{\alpha\textbf{k}}}{V}\right )^2 \delta(\omega - \epsilon_{\alpha\textbf{k}}) ~
= ~ -\frac{1}{\pi} \text{Im}~G_{0\sigma}^{(0)}(\omega)
\end{equation} 
where $G_{0\sigma}^{(0)}(\omega)\equiv \langle \langle f_{0\sigma}^{\phantom{\dagger}} ; f_{0\sigma}^{\dagger} \rangle \rangle_{\omega}^{(0)}$ is the free Green function for the local bath site,  $f_{0\sigma}$ (correlation functions of the type $\langle \langle \hat{A} ; \hat{B} \rangle \rangle_{\omega}$ are simply 
the Fourier transform of the usual retarded functions $-i\theta(t_1-t_2)\langle \{ \hat{A}(t_1),\hat{B}(t_2) \} \rangle$).

The conduction electron and hybridization terms of the Hamiltonian enter in such quantum impurity problems through a single `hybridization function',\cite{hewson} given generally by
\begin{equation}\label{eq:hybfunc}
\Gamma_{\sigma}(\omega) = -V^2 \text{Im}~G_{0\sigma}^{(0)}(\omega) \equiv \Gamma^R_{\sigma}(\omega) - i\Gamma_{\sigma}^{I}(\omega) \;,
\end{equation}
with $\Gamma_{\sigma}^{I}(\omega)=\pi V^2 \rho_{\sigma}^{(0)}(\omega)$, and where the real part, $\Gamma^R_{\sigma}(\omega)$, follows by Hilbert transformation of the imaginary part, $\Gamma_{\sigma}^{I}(\omega)$. As such, the free DOS and the tunnel-coupling $V$ specify completely the effect of the leads on the TQD. For simplicity, we take a wide flat DOS, which is the conventional and generic case\cite{hewson} relevant to most metallic hosts: $\rho_{\sigma}^{(0)}(\omega)=\tfrac{1}{2D} \theta(D-|\omega|)$, such that 
$\Gamma_{\sigma}^{I}(\omega) \equiv \Gamma = \pi \rho V^2$ is defined inside the band, $-D\le \omega \le D$, and $\rho=1/(2D)$.


\subsection{Numerical Renormalization Group}\label{sec:nrg}
 
The Numerical Renormalization Group\cite{wilson} (NRG) is a non-perturbative technique for treating quantum impurity problems, such as the lead-coupled TQD system (for a recent review, see Ref.~\onlinecite{nrg:rev}). Numerically exact thermodynamic and dynamic quantities can be calculated with NRG on essentially all relevant energy scales --- from the conduction electron bandwidth $D$ (typically a few eV in bulk metals), down to the Kondo scale $T_K$ (which is an exponentially-small\cite{hewson} fraction of $D$). 

The first step\cite{wilson} is to divide the conduction electron DOS into intervals whose width decreases exponentially (the points separating intervals are $\omega_n=\pm D\Lambda^{-n}$, with $\Lambda>1$ and $n=0,1,2 ... $). The spectrum is then discretized by replacing the DOS in each interval by a single pole of the same total weight. The `Wilson chain'\cite{wilson,kww} is then defined by writing the conduction electron Hamiltonian as a 1d tight-binding chain of form Eq.~(\ref{eq:Hleads2}), and parameters ${h_n}$, ${e_n}$ chosen so that the free DOS at the TQD site corresponds to the discretized spectrum. 
This discretized model is then diagonalized iteratively, with high-lying states discarded at each step (the truncation being justified because the strength of the hoppings $h_n\sim \Lambda^{-n/2}$ decrease exponentially down the chain\cite{wilson,kww}). 

One thus examines the behavior of the system on progressively lower energy scales, and thermodynamics can be built up as a function of temperature.\cite{wilson,nrg:rev,kww} Dynamical quantities, such as spectral functions, can also be obtained  
from the full density matrix,\cite{asbasis,fdmnrg} calculated iteratively in the Anders-Schiller basis.\cite{asbasisprl}
For the calculations presented in this paper, we use a discretization parameter $\Lambda=3$, and retain around $3000$ states at each iteration.


\subsection{Physical quantities}\label{sec:quantities}

We consider below the contribution to thermodynamics arising from the lead-coupled TQD system, specifically the so-called excess quantities\cite{wilson,kww} 
$\langle \hat{\Omega} \rangle_{\text{imp}} = \langle \hat{\Omega} \rangle - \langle \hat{\Omega} \rangle_{0}$, with $\langle \hat{\Omega} \rangle_{0}$ denoting the thermal average in the absence of the TQD itself. We focus on the entropy $S_{\text{imp}}(T)$, uniform spin susceptibility $\chi_{\text{imp}}(T)=\langle (\hat{S}^z)^2\rangle_{\text{imp}}/T$, and `excess' charge $n_{\text{imp}}=\langle \hat{Q} \rangle_{\text{imp}}$ (here $\hat{S}^z$ and $\hat{Q}$ refer to the spin and charge of the entire system). These quantities show characteristic  signatures of the underlying fixed points, reached under renormalization on progressive reduction of the temperature/energy scale. The renormalization group flow between such fixed points shows up as crossover behavior in thermodynamics, which thus allows determination of emergent energy scales in the problem, such as the Kondo temperature, $T_K$.

\begin{figure*}
\begin{center}
\includegraphics[height=75mm]{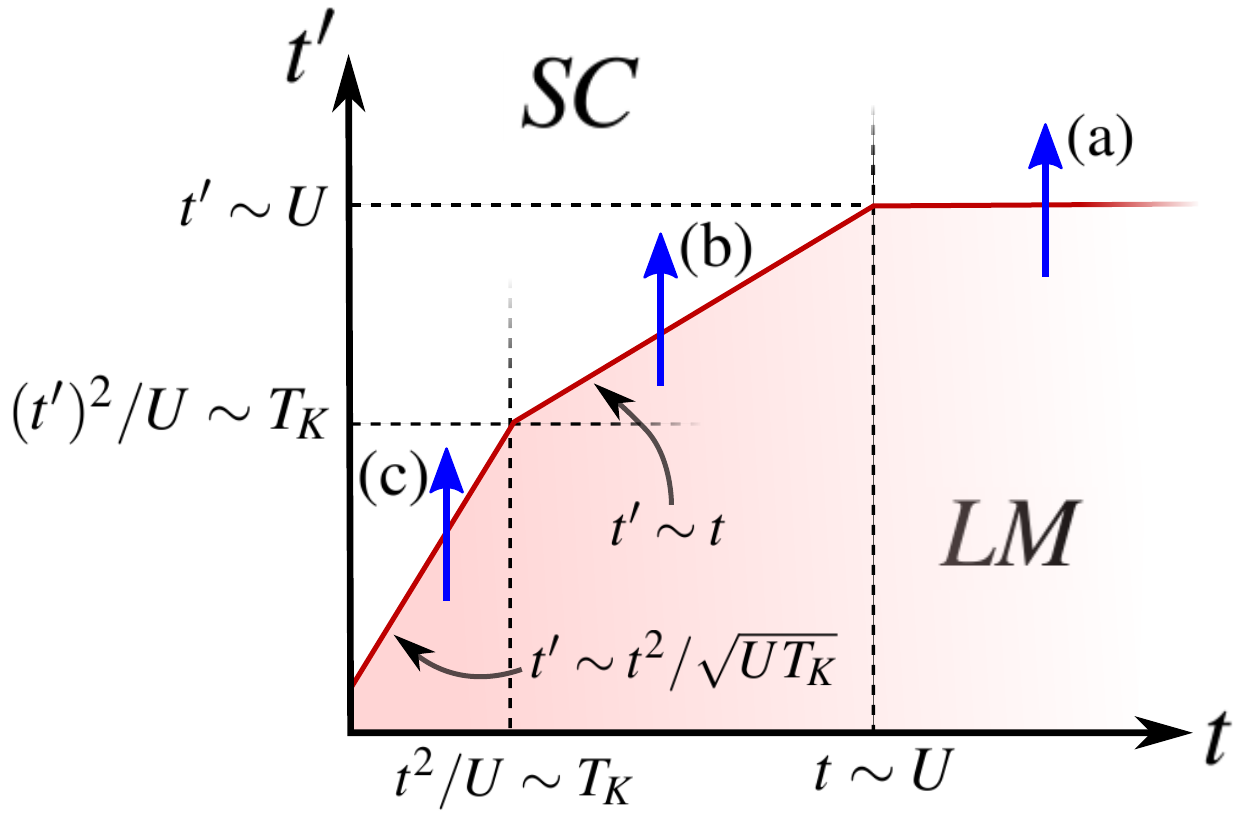}
\caption{\label{fig:pd} Schematic phase diagram for the TQD in the parity-symmetric case, with $\epsilon =-U/2$. For a given $U$ ($\gg \Gamma$), both local moment (LM) and Kondo strong coupling (SC) phases can be accessed on tuning the interdot tunnel-couplings $t$ and $t'$. Regimes and crossovers discussed in the text. The QPT between LM and SC phases indicated by arrows (a)--(c) is considered explicitly in the following sections.}
\end{center}
\end{figure*}

We also consider dynamical properties --- in particular dot spectral functions (or, loosely, local densities of states), given by $D_{i\sigma}(\omega) = -\tfrac{1}{\pi} \text{Im}~G_{ii,\sigma}(\omega)$ in terms of the local Green function for that dot, $G_{ii,\sigma}(\omega)=\langle \langle d_{i\sigma}^{\phantom{\dagger}} ; d_{i\sigma}^{\dagger} \rangle \rangle_{\omega}$.
In the zero-bias (equilibrium) limit of interest, it is naturally the local spectrum of dot 1 (itself directly tunnel-coupled to the leads) which determines the
differential conductance $G_{c}(T)$ between source and drain leads, mediated via the TQD. This is given exactly\cite{meir} by
\begin{equation}\label{mw}
\frac{G_c(T)}{G_0} = \int_{-\infty}^{\infty} -\frac{\partial f(\omega)}{\partial \omega}~ \pi \Gamma D_{1\sigma}(\omega,T)~d\omega \;
\end{equation}
where $f(\omega)=[1+\exp(\omega/T)]^{-1}$ is the Fermi function (and $\omega =0$ is the Fermi level). Here, $G_0=2e^2h^{-1}\tilde{G}_0$,
where the dimensionless quantity  $\tilde{G}_0=4\Gamma_s\Gamma_d/(\Gamma_s+\Gamma_d)^2$ embodies simply the relative coupling strength of dot 1 to source and drain leads; such that $G_0=2e^2h^{-1}$ is maximal in the symmetric case $\Gamma_s=\Gamma_d$. 
At zero temperature, the conductance is thus controlled by the behavior of the spectrum at the Fermi level, 
\begin{equation}\label{T0cond}
\frac{G_c(0)}{G_0} = \pi\Gamma D_{1\sigma}(\omega=0,T=0) = \sin^2(\delta) \;,
\end{equation}
where the final identity follows from the definition of the scattering phase shift $\delta=\arg[G_{1\sigma}(\omega=0,T=0)]$, together with the fact that all electron scattering vanishes at the Fermi level.


\section{Results: symmetric case}\label{sec:parity}

We consider first the full TQD model, Eq.~(\ref{eq:H}), in the parity-symmetric case $t_{12}=t_{13}\equiv t$ and $t_{23}\equiv t'$ (see Fig.~\ref{fig:tqd}). 
In this limit the Hamiltonian is invariant to swapping the dot labels $2\leftrightarrow 3$, meaning formally that $[H,\hat{P}_{23}]=0$ for the permutation operator $\hat{P}_{23}$. All states can thus be classified according to parity $P_{23}=\pm 1$ (since $\hat{P}_{23}^2=1$). As shown in Ref.~\onlinecite{akm:tqd1ch} this allows for the possibility of a (zero temperature) quantum phase transition (QPT) between two parity-distinct phases, which are of Kondo screened and local moment types. That scenario is explored in detail below, focusing primarily on the strongly correlated case 
$U\gg \Gamma$. Since the most interesting physical behavior arises when the TQD is in essence triply occupied, we consider the representative case 
$\epsilon=-U/2$.
The largest energy scale of the problem, the bandwidth $D$, is taken as $D=100\Gamma$.


\subsection{Phase diagram}\label{sec:pd}

Before discussing specific results for the symmetric TQD, in Fig.~\ref{fig:pd} we present the phase diagram which highlights the relevant regimes, crossover scales and phases, verified directly by full NRG calculations. The parity-symmetric TQD model supports two phases: a Kondo screened phase with strong coupling (SC) ground state, and a local moment (LM) phase with a doubly-degenerate ground state.\cite{akm:tqd1ch}

As shown below, however, the effective spins which remain free in the LM phase depend on the underlying parameters in the 
$(t, t^{\prime})$-plane. Likewise, Kondo screening proceeds by different mechanisms in the SC phase depending on relative magnitudes of the interdot tunnel-couplings. 
When the interdot tunnel-couplings are large compared to the interaction, $t\gg U$ (regime (a) of Fig.~\ref{fig:pd}), 
the relevant TQD states are simple molecular orbitals (MOs); with the lowest doublet state remaining free down to $T=0$ when $t' \lesssim U$. 
By contrast, for $t, t' \ll U$ each dot is essentially singly-occupied. In this case, dot 1 (connected directly to the leads) can undergo the Kondo effect on the scale of $T_K$ --- provided the interdot tunnel-couplings are weak ($t^2/U\ll T_K$). The strength of the effective coupling between the remaining dots 2 and 3 then determines the ultimate ground state of the system, with a level-crossing QPT occurring between LM and SC states as $t'$ is increased through $t'\sim t^{2}/\sqrt{U T_K}$ (regime (c) of Fig.~\ref{fig:pd}). 
If by contrast the interdot couplings are stronger, $t^2/U\gg T_K$, the lowest-energy TQD doublet state has weight on all three dots. This results -- regime (b) of Fig.~\ref{fig:pd} --
in an effective antiferromagnetic coupling to the leads if $t'\gtrsim t$, in which case the Kondo effect is operative and the SC phase obtains; or, if $t' \lesssim t$, in an effective ferromagnetic coupling to the leads, such that the TQD doublet remains free down to $T=0$ in this LM phase.

In the following, we discuss in some detail the behavior in these three  distinct regimes.


\subsection{Molecular orbital regime}\label{sec:mo}

We examine first the simplest case, in which the interdot tunnel-couplings are strong, $t\gg U$. Here, the isolated TQD states are essentially non-interacting MOs. To obtain a 
handle on the problem in this regime, we thus derive a low-energy effective model for the non-interacting TQD ($U=0$), then incorporate the effect of interactions within lowest-order perturbation theory in $U$. 

When $U=0$, a simple canonical transformation of the dot operators 
$\{d_{1\sigma}, d_{2\sigma}, d_{3\sigma} \} \rightarrow \{d_{u\sigma}, d_{l\sigma}, d_{o\sigma} \}$ brings $H_{\text{TQD}}$ in Eq.~\ref{eq:H} to diagonal form,
\begin{equation}\label{eq:u0}
H_{\text{TQD}}^{U=0} = \sum_{\sigma}(E_{u}\hat{n}_{u\sigma}+E_{o}\hat{n}_{o\sigma}+E_{l}\hat{n}_{l\sigma}),
\end{equation}
where $\hat{n}_{\alpha \sigma} = d^{\dagger}_{\alpha \sigma} d_{\alpha\sigma}$ ($\alpha = u,l,o$).
The single-particle levels (MO energies) are given by
$E_{o} =  \epsilon - t'$ and $E_{u/l} =   \epsilon + \tfrac{1}{2} t' \pm \tfrac{1}{2} \sqrt{8t^{2}+t'^{2}}$, and
for $t^{\prime}<t$ have the relative ordering $E_{l} <E_{o} <0<E_{u}$. 
The difference in energy between the 4-electron state and the 3-electron state is thus 
$E_{\Delta} \equiv E_{o} = \epsilon-t' <0$. The lowest many-particle TQD state is thus always a 4-electron state when $U=0$. 

Intuition suggests however that the Coulomb repulsion will favor a 3-electron state.
Treating perturbatively the interaction part of the Hamiltonian, $H_{I}=U\sum_i \hat{n}_{i\uparrow}\hat{n}_{i\downarrow}$, we thus calculate the correction to the isolated TQD states to first-order in $U$. The resultant energy difference between the 4- and 3-electron states is then $E_{\Delta}' = \epsilon -t' +\tfrac{1}{2} U (1+\gamma^2) +\mathcal{O}(U^2)$, with $2\gamma^2=1-[1+8(t/t')^2]^{-1/2}$. Hence for $\epsilon=-U/2$ as considered explicitly, $E_{\Delta}' =U/4-t'$, indicating that the 3-electron TQD state is indeed favored when $t^{\prime} \lesssim U/4$.

Of course, the interesting behavior arises on coupling the TQD to the leads. In the relevant case of $\Gamma \ll U$ ($\ll t$), one can obtain an effective model valid at low temperatures/energies $\ll E_{\Delta}'$, by projecting the full Hamiltonian onto the 4-electron TQD state (for $E_{\Delta}'<0$) or the 3-electron state (for $E_{\Delta}'>0$). This is achieved by a Schrieffer-Wolff transformation (SWT)\cite{sw}
to second order in $H_{\text{hyb}}$, eliminating perturbatively excitations to higher-lying TQD states.

In the effective 4-electron sector, the spin-singlet TQD state essentially decouples from the leads. This `frozen impurity' fixed point is continuously connected to the Kondo screened ground states\cite{hewson,kww} (discussed further below). More interestingly, in the 3-electron sector of the TQD, one obtains via a SWT the effective model 
\begin{equation}\label{eq:effKondolarget}
H^{\textit{eff}}= H_{\text{leads}} + J^{\ddagger}_{\mathrm{K}} \hat{\mathbf{S}}\cdot \hat{\mathbf{s}}(0) ,
\end{equation}
with $\hat{\mathbf{S}}$ a spin-$\tfrac{1}{2}$ operator for the lowest (doublet) 3-electron TQD state, and $\hat{\mathbf{s}}(0)$ the spin density of conduction electrons at the TQD. Eq.~\ref{eq:effKondolarget} is a model of Kondo form,\cite{hewson} with its underlying physics well known to depend crucially on the sign of the exchange coupling, $J^{\ddagger}_{\mathrm{K}}$. 
Importantly, we find here $J^{\ddagger}_{\mathrm{K}}<0$, wherein the \emph{ferromagnetic} Kondo effect arises.\cite{hewson,lmcor1,lmcor2} Kondo quenching of the TQD spin-$\tfrac{1}{2}$ is thus inoperative, with the TQD decoupling from the leads on renormalization under reduction of the temperature/energy scale, leaving asymptotically a free local moment as $T\rightarrow 0$. The residual entropy at the corresponding LM fixed point is thus $S_{\text{imp}}=\ln(2)$, and leading irrelevant corrections to the fixed point are non-analytic.\cite{lmcor1,lmcor2} 

\begin{figure}
\begin{center}
\includegraphics[height=60mm]{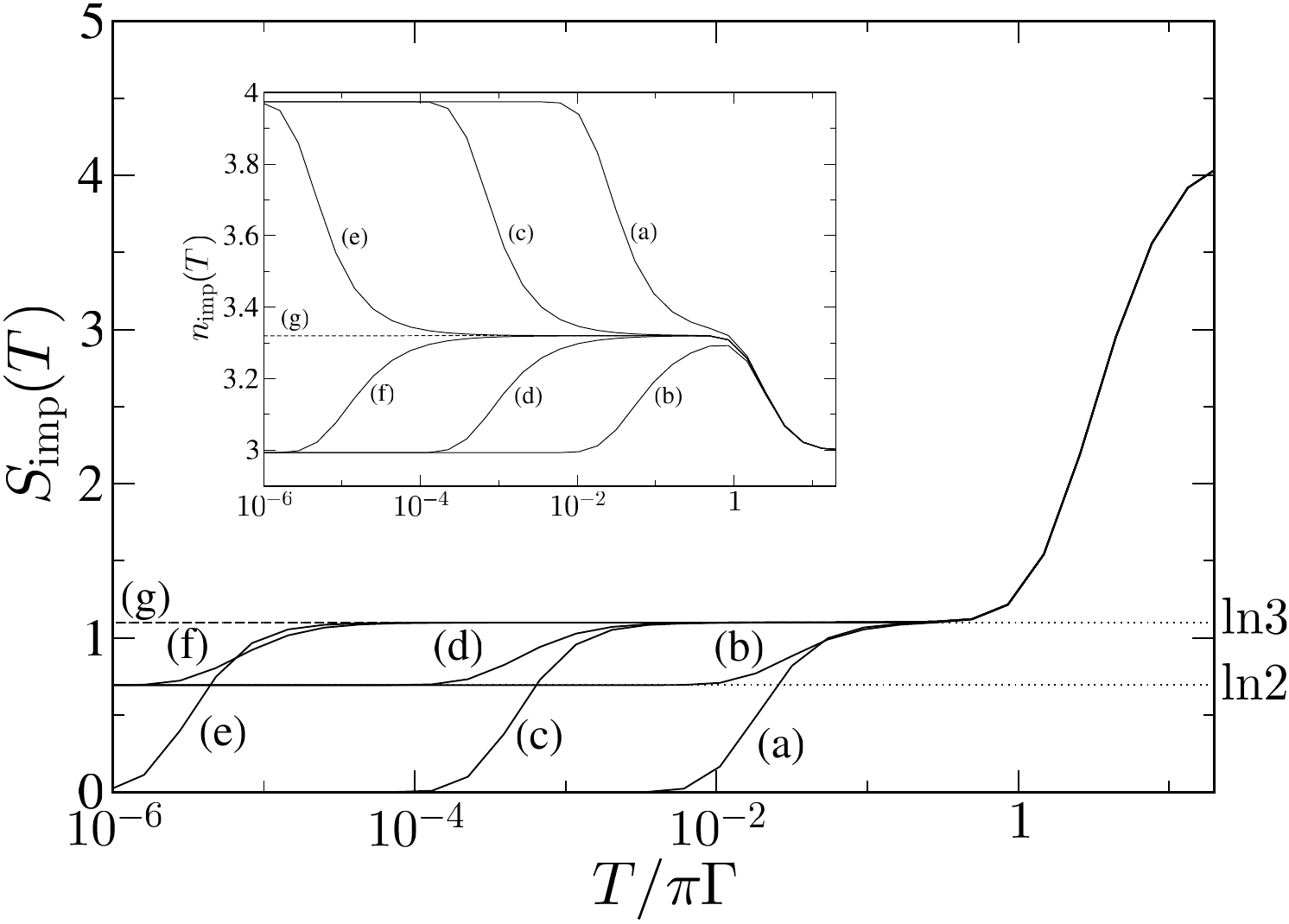}
\caption{\label{fig:mosym} Entropy $S_{\mathrm{imp}}(T)$ vs $T/\pi \Gamma$ for systems in the MO regime. Plotted for common $U/\pi\Gamma=10$ and $t/\pi\Gamma=5$, with $t'/\pi\Gamma$ tuned to approach the transition [line (g), $t_c'/\pi\Gamma \approx 2.45793$] from the SC phase [$t'/\pi\Gamma=2.5$, $2.459$ and $2.45794$ for lines (a), (c) and (e)] and from the LM phase [$t'/\pi\Gamma=2.4$, $2.457$ and $2.45792$ for lines (b), (d) and (f)]. The inset shows the excess charge, $n_{\text{imp}}(T)$, for the same systems.}
\end{center}
\end{figure}

Since the ground states for $E_{\Delta}'>0$ (LM) and $E_{\Delta}'<0$ (SC) \emph{cannot} be continuously connected, a QPT is expected around $t'\sim U/4$ [see arrow (a) of Fig.~\ref{fig:pd}].
This physical picture is confirmed directly in Fig.~\ref{fig:mosym}, where  NRG results for the full TQD model are presented. Specifically, we show the TQD contribution to the entropy $S_{\text{imp}}(T)$ as a function of temperature, for systems approaching the QPT between LM and SC phases. For $T \gg |E_{\Delta}'|$, the 3- and 4- electron states are quasi-degenerate, and a $\ln(3)$ entropy thus results. Flow to either the LM ground state with $S_{\text{imp}}=\ln(2)$ or the SC ground state with $S_{\text{imp}}=0$ occurs on the scale of $T_{\Delta} \equiv |E_{\Delta}'|$, which as such vanishes linearly with $|t'-t'_c|$ as the transition at the critical $t'_c\simeq U/4$ is approached from either side (and which linearity is symptomatic of a parity-protected \emph{level-crossing} 
QPT~\cite{akm:tqd1ch}).

The inset shows the excess charge\cite{hewson,kww} due to the TQD, $n_{\text{imp}}(T)$, for the same systems; confirming that the transition at $T=0$ occurs between effective 3- and 4-electron TQD states.
These results also show that precisely at the QPT (curve (g) in Fig.~\ref{fig:mosym}), the problem can be understood in terms of coexisting SC and LM fixed points -- for example the $T=0$ charge 
$n_{\mathrm{imp}} = 3\tfrac{1}{3} = \tfrac{1}{2+1}( 3\times 2 + 4\times 1)$, 
is simply an average of the triple charge characteristic of the doubly-degenerate LM fixed point and the quadruple charge of the singly-degenerate SC fixed point.


\subsection{Spin regime: strong interimpurity coupling}\label{sec:bigJ}

More subtle behavior arises in the strongly-correlated case $U\gg t,t',\Gamma$, where 
each dot is essentially singly-occupied for temperatures $T\ll U$. However, the tunnel-couplings do generate effective \emph{exchange} couplings between the dots. This can be seen by projecting (SWT) the full $H_{TQD}$ onto the singly-occupied manifold of TQD states, perturbatively eliminating virtual excitations to 2- and 4-electron TQD states to second-order in the tunnel-couplings $t$, $t'$ and $V$. The resulting effective spin model is then given by
\begin{equation}\label{eq:spinmodel}
H_{\mathrm{spin}} = H_{\mathrm{leads}} +J \hat{\mathbf{S}}^{}_{1} \cdot (\hat{\mathbf{S}}^{}_{2}+\hat{\mathbf{S}}^{}_{3}) +J'\hat{\mathbf{S}}^{}_{2}\cdot \hat{\mathbf{S}}^{}_{3} +J^{}_{\mathrm{K}}\hat{\mathbf{S}}^{}_{1}\cdot \hat{\mathbf{s}}(0)
\end{equation}
where $\hat{\mathbf{S}}^{}_{i}$ is a spin-$\tfrac{1}{2}$ operator for dot $i$, $\hat{\mathbf{s}}(0)$ is the conduction electron spin density at the TQD as before, and  the exchange couplings are given by
\begin{equation}\label{eq:exchanges}
J = \frac{4 t^{2}}{U} \qquad ; \qquad J' =  \frac{4 t'^{2}}{U} \qquad ; \qquad 
\rho J_{\mathrm{K}} =  \frac{8\Gamma}{\pi U} \;.
\end{equation}

Two parity-distinct doublet ground states are then obtained for the isolated TQD, depending on whether $t'>t$ (even parity) or $t'<t$ (odd parity). 
In Ref.~\onlinecite{akm:tqd1ch} we derived analytically an effective low-energy model to describe this lead-coupled TQD system for $t^2/U \gg T_K$ but $t\ll U$, corresponding to regime (b) in Fig.~\ref{fig:pd}. In either case, $t'\gg t$ or $t'\ll t$, the effective low-energy model is again a Kondo model  Eq.~\ref{eq:effKondolarget}, describing the residual coupling of the lowest TQD doublet state to the leads. Importantly however, $J^{\ddagger}_{\mathrm{K}}=-\tfrac{1}{3}J_K$ for $t'\ll t$ and $J^{\ddagger}_{\mathrm{K}}=+J_K$ for $t'\gg t$.
The ferromagnetic Kondo effect\cite{hewson} thus again arises for $t'<t'_c \simeq t$, with the lowest TQD doublet decoupling asymptotically at low energies. However, for  $t'>t'_c$ the regular spin-$\tfrac{1}{2}$ antiferromagnetic Kondo effect\cite{hewson} drives the system to a strong coupling state below the Kondo scale $T_K \sim D \exp(-1/\rho J_K)$, in which all TQD degrees of freedom are quenched by the lead conduction electrons. The entropy is thus $S_{\text{imp}}(0)=0$, corresponding to the Kondo singlet ground state.\cite{hewson,wilson,kww} On tuning $t'$ through $t'_c\simeq t$, a level-crossing QPT occurs between LM and SC phases.\cite{akm:tqd1ch}

\begin{figure}
\begin{center}
\includegraphics[height=50mm]{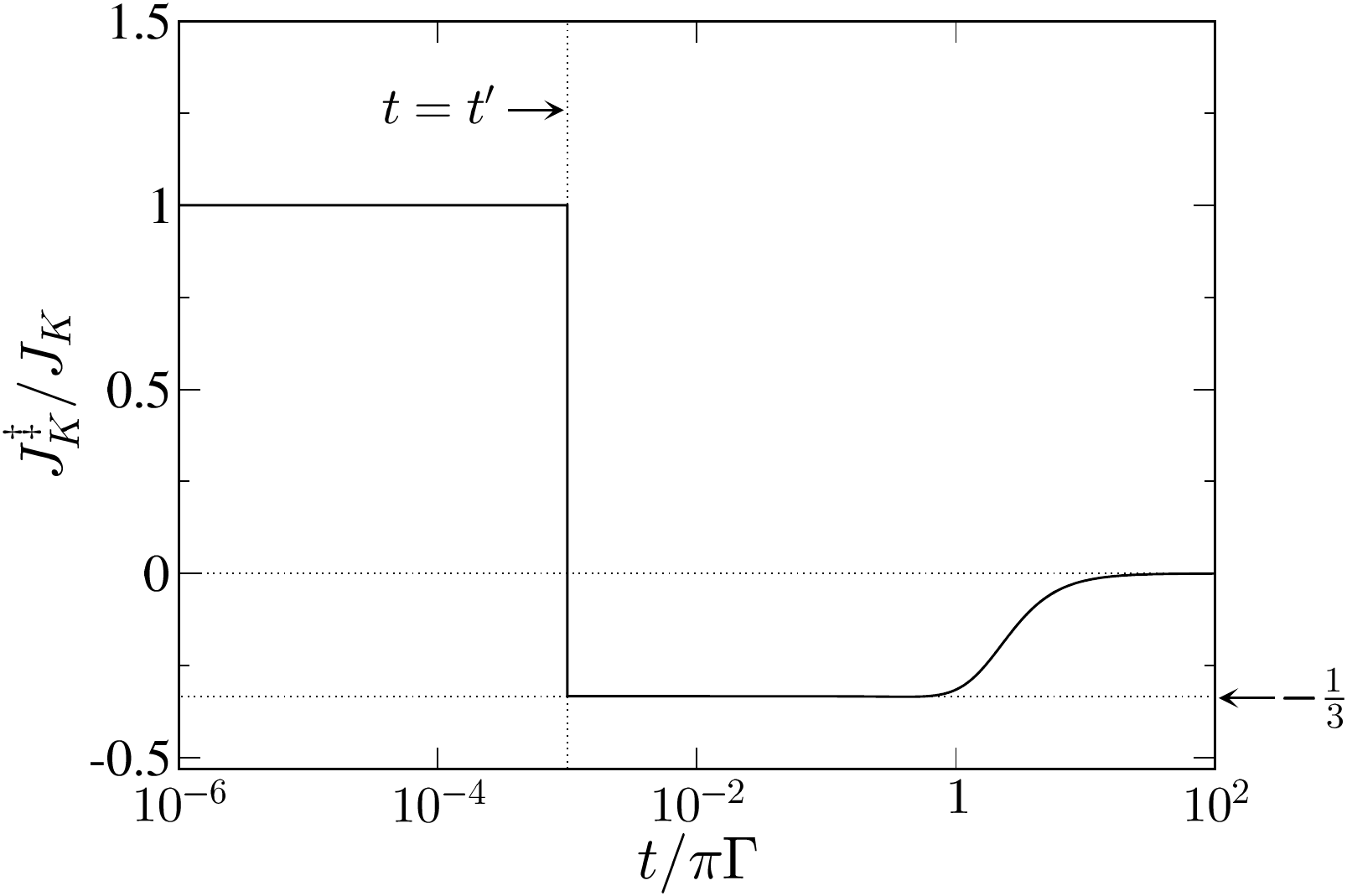}
\caption{\label{fig:largeJsym} Effective coupling $J_K^{\ddagger}/J_K$ vs $t/\pi\Gamma$ for the low-energy Kondo model Eq.~\ref{eq:effKondolarget} in the regime $t, t' \ll U$. Here $t'/\pi\Gamma=10^{-3}$ and $U/\pi\Gamma=8$.}
\end{center}
\end{figure}

One can in fact derive more generally an effective low-energy Kondo model of form Eq.~\ref{eq:effKondolarget}, working directly with $H_{TQD}$ (i.e.\ without first projecting into the spin sector, as above). The effective Kondo model is obtained simply by projecting the full Hamiltonian onto the reduced TQD Hilbert space spanned by the lowest eigenstate of the numerically-diagonalized $H_{TQD}$. For any $t' < t'_{c} \simeq t \ll U $ the odd-parity TQD doublet couples to the leads, while the even parity doublet couples for $t_{c}' < t' \ll U$. The QPT can of course also be realized by tuning $t$, keeping $t'$ fixed (Fig.~\ref{fig:pd}), and a numerical calculation of the effective coupling so obtained, $J^{\ddagger}_{\mathrm{K}}$, is shown in Fig.~\ref{fig:largeJsym}. The asymptotic values\cite{akm:tqd1ch} $J^{\ddagger}_{\mathrm{K}}/J_{K}=-\tfrac{1}{3}$ and $+1$ in the LM and SC phases are recovered in the singly-occupied limit; and the discontinuous change in the sign of $J^{\ddagger}_{\mathrm{K}}$ is seen clearly at the point $t'=t$, due to the level crossing of TQD states. NRG calculations on the full model (without any low-energy projections) indeed confirm that the transition occurs very close to $t'_c = t$.

\subsection{Spin regime: weak interimpurity coupling}\label{sec:smallJ}

As suggested by arrow (c) in the phase diagram Fig.~\ref{fig:pd}, a QPT between LM and SC phases arises even when the interdot couplings $t$ and $t'$ are very small. In this case however, the mechanism for local moment formation and Kondo screening is rather subtle.

Since $t$, $t'$, $\Gamma \ll U$, the dots are still essentially singly-occupied, so the effective spin Hamiltonian Eq.~\ref{eq:spinmodel} remains valid. In the limit where $J=0$ (arising when $t=0$), dot 1 decouples from dots 2 and 3, and 
undergoes the regular Kondo effect with the lead to which it is coupled, being screened below $T\sim T_K$ (with $T_K$ the Kondo scale $T_K\sim D \exp(-1/\rho J_K)$ as above).\cite{hewson} This must also in fact remain the case for small but non-vanishing $J\ll T_K$, since the quenched Kondo singlet is already formed on the scale of $T_{K}$.
However, this fixed point contains residual unquenched degrees of freedom corresponding to the spins on dots 2 and 3. In order to analyze the stability of this fixed point, one must thus determine whether there is an effective  coupling of these degrees of freedom to the remaining Fermi liquid bath states of the lead.

The effective exchange coupling acting directly between dots 2 and 3 is given simply by $J'=4t'^2/U$ (see Eqs.~\ref{eq:spinmodel}, \ref{eq:exchanges}), such that the local $S=0$ singlet state is lower in energy than the local $S=1$ triplet by $E_{\Delta}=E_T-E_S=J'$. However, there is an additional effective coupling between dots 2 and 3 due to an RKKY-type interaction mediated by the Kondo singlet formed between dot 1 and the leads.
Virtual polarization of this Kondo singlet is readily shown to generate an effective \emph{ferromagnetic} contribution to the coupling between the spins of dots 2 and 3;  with second-order perturbation theory in the coupling $J$ within a Wilson chain formalism\cite{akm:oddimp,akm:2ckin2ik} thereby yielding a renormalized singlet-triplet splitting, $E_{\Delta}'=J'-\lambda J^2/T_K$ (with $\lambda=\mathcal{O}(1)$ an undetermined positive constant).
When $E_{\Delta}'>0$, the 2-3 singlet thus lies lowest and decouples from the rest of the system. In consequence, at temperatures $T\ll |E_{\Delta}'|$ the \emph{entire} system is in a singlet state, characteristic of the SC phase of regime (c) in Fig.~\ref{fig:pd}.

By contrast, when $E_{\Delta}'<0$ the \emph{triplet} formed between dots 2 and 3 lies lowest in energy. Thus in this case there are still residual TQD degrees of freedom at temperatures $T\ll |E_{\Delta}'|$. To determine the stability of this state, we must again consider the effective coupling between this triplet and the rest of the system. Within the Wilson chain picture, it can be shown that the TQD triplet experiences an effective coupling to the remaining Fermi liquid bath states of the lead, mediated via the Kondo singlet formed with dot 1. The mechanism here is in fact completely analogous to that occurring in odd quantum dot chains with weak interdot coupling;\cite{akm:oddimp} or in the asymmetric two-impurity Kondo problem.\cite{akm:2ckin2ik} For $E_{\Delta}'<0$, the effective low-energy model follows as
\begin{equation}\label{eq:s1model}
\hat{H}^{\textit{eff}}_{S=1}=\tilde{H}_{\text{leads}}+J^{*}_{\mathrm{K}}\mathbf{\hat{S}}\cdot \mathbf{\tilde{s}}(0) \;,
\end{equation}
where $\mathbf{\hat{S}}$ is now a spin-1 operator for the residual TQD state, $\tilde{H}_{\text{leads}}$ is the free conduction electron Hamiltonian with the Wilson chain `zero orbital' removed due to the first-stage Kondo screening involving dot 1; and $\mathbf{\tilde{s}}(0)$ is its spin density at the TQD. 
The key result is that the effective coupling $\rho J^{*}_{\mathrm{K}}\sim t^2/T_{K}^{2} >0$ is now \emph{antiferromagnetic}, and the effective low-energy model Eq.~\ref{eq:s1model} is thus a realization of the famous single-channel spin-1 Kondo model of Nozi\`eres and Blandin \cite{nozieres}.

The rich physical behavior of the spin-1 Kondo model\cite{nozieres,lmcor1,lmcor2,akm:tqd1ch,CW_multilevel,CW_multilevel2} is thus expected on the lowest energy scales. The antiferromagnetic coupling,  $J^{*}_{\mathrm{K}}$ is renormalized upward on reduction of the temperature scale, resulting in quenching (or `underscreening') of the $S=1$ `impurity' below $T_K^* \sim D \exp(-1/\rho J^{*}_{\mathrm{K}})$ ($\ll |E_{\Delta}'| \ll T_K$)~\cite{nozieres}; such that the 
ground state comprises a residual free spin-$\tfrac{1}{2}$ local moment, with leading RG-irrelevant corrections to the fixed point that are ferromagnetic (and non-analytic).\cite{lmcor1,lmcor2}

In consequence, there is again a QPT on tuning $t'$, separating a Kondo screened SC phase from a LM phase (regime (c) of Fig.~\ref{fig:pd}). The transition itself is expected to occur at $E_{\Delta}' = J' -\lambda J^{2}/T_{K} \simeq 0$, implying $J' \sim J^2/T_K$  [or equivalently $t' \sim t^2/\sqrt{U T_K}$, see Fig.~\ref{fig:pd}, arrow (c)].

\begin{figure}
\begin{center}
\includegraphics[height=60mm]{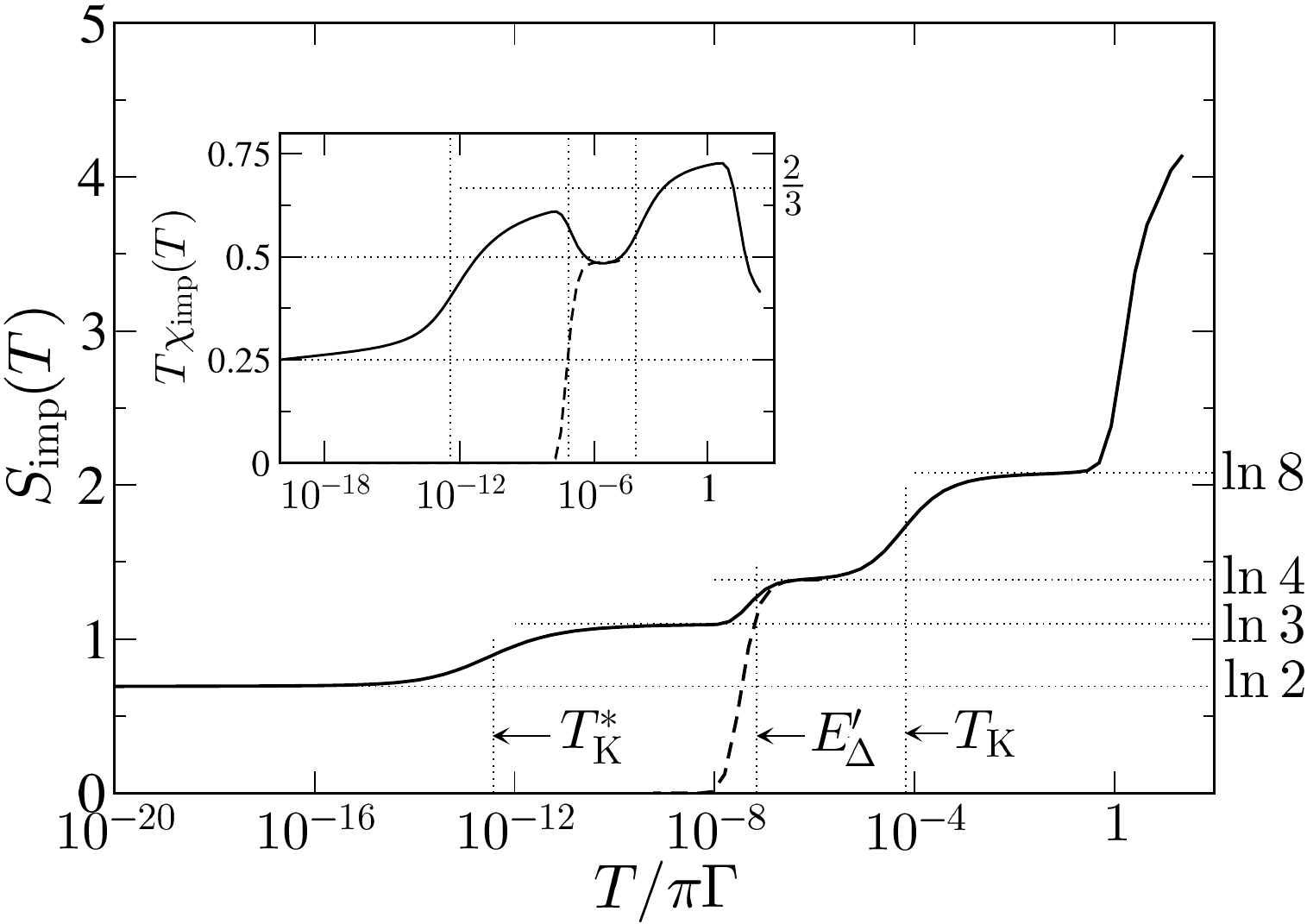}
\caption{\label{fig:smallJsym_td} 
Thermodynamics for the case of small interdot tunnel-couplings, $t^2/U\ll T_K$. The main panel shows the TQD contribution to entropy $S_{\text{imp}}(T)$ vs $T/\pi\Gamma$, while the inset shows $T\chi_{\text{imp}}(T)$. Solid line for a system in the LM phase ($E_{\Delta}'<0$) ; dashed line in the SC phase ($E_{\Delta}'>0$). $U/\pi\Gamma=7$, $t/\pi\Gamma=6\times 10^{-3}$ and $t'/\pi\Gamma=1\times 10^{-3}$ (LM) or $6\times 10^{-3}$ (SC).
}
\end{center}
\end{figure}

To illustrate this complex behavior, Fig.~\ref{fig:smallJsym_td} shows full NRG results for the TQD contribution to entropy and magnetic susceptibility for systems close to the transition. Both systems have $T_K/\pi\Gamma \approx 10^{-4}$ and similar $|E_{\Delta}'|/\pi\Gamma \approx 10^{-7}$. The solid line corresponds to the LM phase, and the dashed line to the SC phase. 
On the temperature scale $T\sim U$, the dots become singly occupied, and so $S_{\text{imp}}=3\ln(2)$ and $T\chi_{\text{imp}}=\tfrac{3}{4}$ corresponding to three free spins-$\tfrac{1}{2}$. As the temperature is reduced below $T_K$, dot 1 is screened by the Kondo effect, leaving two quasi-degenerate spins on dots 2 and 3 (yielding thereby $S_{\text{imp}}=2\ln(2)$ and $T\chi_{\text{imp}}=\tfrac{2}{4}$). For $E_{\Delta}'>0$ (dashed line), the residual TQD singlet state is lowest, and so $S_{\text{imp}}=0$ and $T\chi_{\text{imp}}=0$ when $T\ll |E_{\Delta}'|$. However, for $E_{\Delta}'<0$ (solid line), the crossover is first to a free residual TQD triplet state on the scale of $|E_{\Delta}'|$; then below $T\sim T_K^*$ to the ultimate stable LM fixed point describing the underscreened $S=1$ Kondo state, with $S_{\text{imp}}=\ln(2)$ and $T\chi_{\text{imp}}=\tfrac{1}{4}$.

\begin{figure}
\begin{center}
\includegraphics[height=60mm]{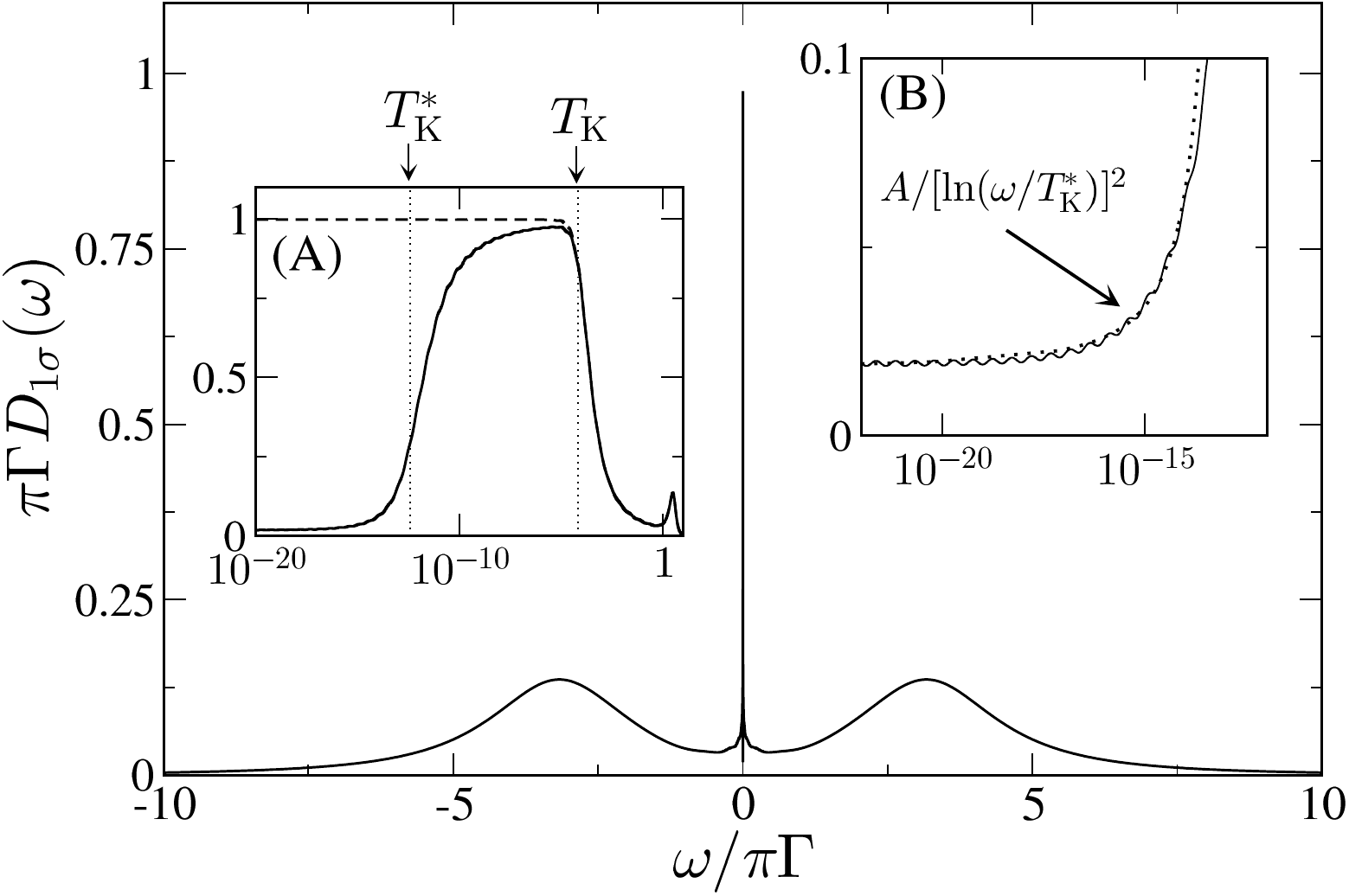}
\caption{\label{fig:smallJsym_spec} Single-particle spectrum of dot 1, $\pi\Gamma D_{1\sigma}(\omega)$ vs $\omega/\pi\Gamma$ at $T=0$, for systems with the same parameters as \ref{fig:smallJsym_td}. Insets (A) and (B) show detail of the low-energy behavior on a logarithmic energy scale. The approach to the LM fixed point is characterized by marginally irrelevant logarithmic corrections, as highlighted in panel (B). However, for $T_K^* \ll |\omega| \ll T_K$ the spectrum approaches the unitarity limit $\pi\Gamma D_{1\sigma}(\omega) = 1$ due to the first-stage Kondo screening of dot 1 [see panel (A)].
}
\end{center}
\end{figure}

Similar behavior is observed in dynamical quantities, such as the $T=0$ spectral function for dot 1, the energy/frequency dependence of which is shown in Fig.~\ref{fig:smallJsym_spec}.
In the SC phase, the classic three-peak structure is observed. At high energies $\omega=\pm U/2$, the only spectral features are the Hubbard satellites, whose origin is simply dot charge fluctuations.\cite{hewson} At low energies $|\omega| \sim T_K$, the narrow Kondo resonance is observed (see in particular inset (A), dashed line); with the unitarity limit $\pi\Gamma D_{1\sigma}(\omega)=1$ reached at the Fermi level, $\omega=0$, being characteristic of the SC fixed point. In the LM phase however (solid lines), first-stage Kondo screening of dot 1 on the scale of $T_K$ is followed by second stage underscreening of the residual TQD triplet state, which as above is mediated via dot 1. This results in a slow crossover on the scale of $T_K^*$ to the LM fixed point, characterized by $\pi\Gamma D_{1\sigma}(\omega) \sim 1/\ln^2|\omega/T_K^*|$ behavior, such that $\pi\Gamma D_{1\sigma}(\omega)=0$ only at $\omega=0$, characteristic of the singular Fermi liquid, and reflecting the marginally irrelevant corrections to the fixed point.\cite{lmcor1,lmcor2,akm:tqd1ch,CW_multilevel,CW_multilevel2}

\begin{figure}
\begin{center}
\includegraphics[height=60mm]{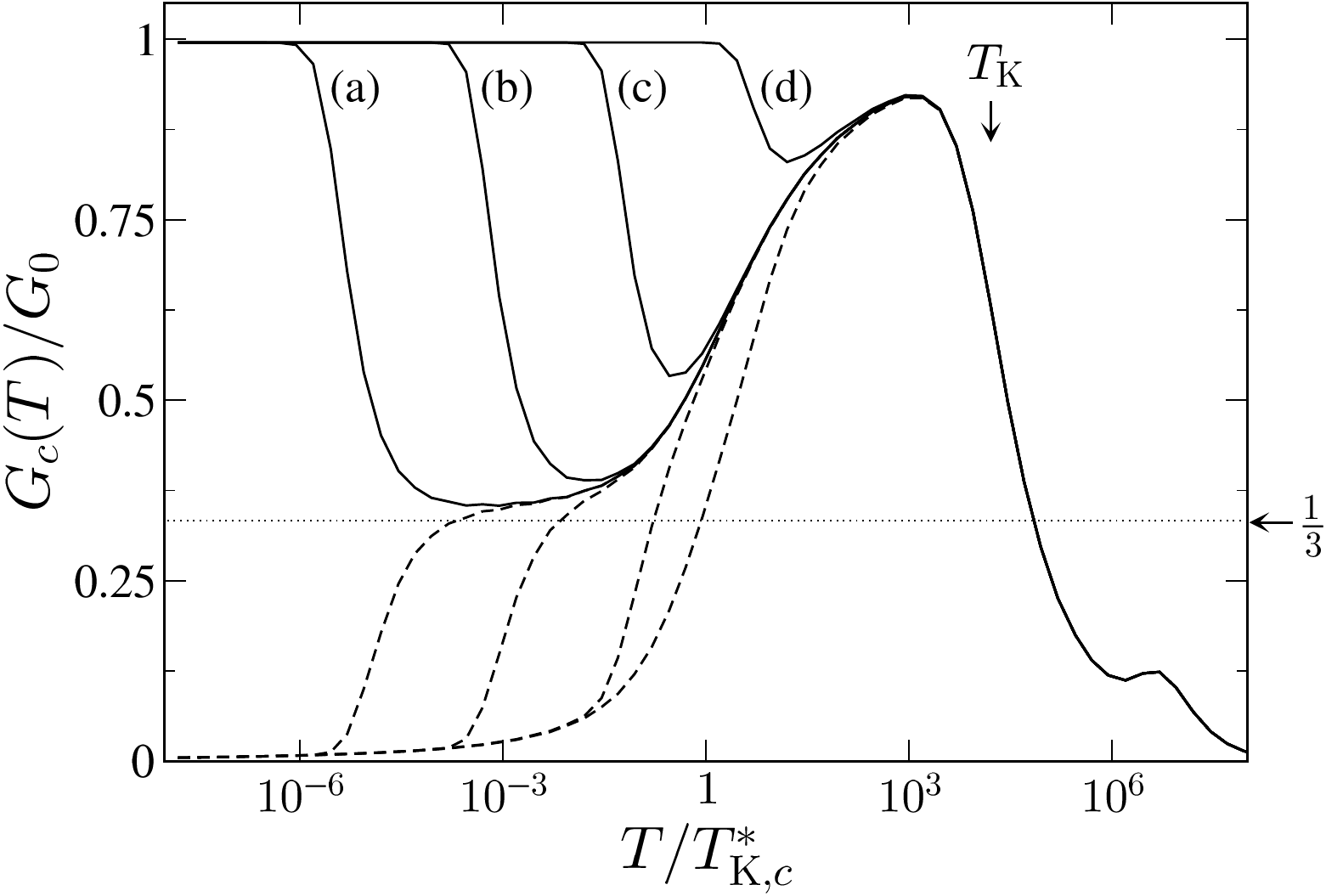}
\caption{\label{fig:smallJsym_cond} Zero-bias conductance through the TQD, $G_c(T)/G_0$ vs $T/T_{K,c}^*$, for systems closely approaching the transition. $T_{K,c}^*$ is the crossover scale upon which the transition fixed point is reached at $t=t_c$. Plotted for $U/\pi\Gamma=4$ and $t'/\pi\Gamma=10^{-2}$, varying $t/\pi\Gamma=t_c\pm\lambda T_K$, with $\lambda = 10^{-7}$, $10^{-5}$, $10^{-3}$ and $10^{-1}$ for lines (a)--(d) respectively. Solid lines for systems in the SC phase; dashed lines for the LM phase. $T_K$ is the common scale for the first-stage Kondo effect involving dot 1, as indicated on the plot. }
\end{center}
\end{figure}

Finally, in Fig.~\ref{fig:smallJsym_cond} we show the zero-bias conductance through the device as a full function of temperature. Solid lines are for $E_{\Delta}'>0$ in the SC phase, approaching progressively closely (from (d)--(a)) the transition at $E_{\Delta}'=0$; while dashed lines are for $E_{\Delta}'<0$ in the LM phase. All systems have a common scale $T_K$, and the data are rescaled in terms of $T/T_{K,c}^*$ (with $T_{K,c}^*$ chosen arbitrarily as the $T$ for which $G_{c}(T)/G_{0} =\tfrac{2}{3}$).
The $\delta=\pi/2$ phase shift in the Kondo screened SC phase\cite{hewson} implies a unitarity $T=0$ conductance $G_c(0)/G_0=1$ from Eq.~\ref{T0cond}; while $\delta=0$ in the LM phase\cite{CW_multilevel} gives vanishing conductance,\cite{CW_multilevel} $G_c(0)/G_0=0$. The behavior at the transition fixed point itself can again be understood here in terms of coexisting LM and SC states; yielding a $T=0$ conductance at the transition of $G_c(0)/G_0=\tfrac{1}{3}$. The full temperature dependence is of course rich, reflecting with decreasing $T$ renormalization group flow to the Kondo SC fixed point on the scale of $T \sim T_K$, followed by flow to the transition fixed point at $T_{K,c}^*$; and finally flow to the stable LM or SC fixed points describing the true ground state, on the scale $T\sim |E_{\Delta}'|$.

\section{Results: distortions}\label{sec:distortions}

We now turn to the more general case of broken parity symmetry, as occurs due to distortions of the triangular TQD structure (Fig.~\ref{fig:tqd}). The presence of  symmetry-breaking perturbations, such as $t_{12}\ne t_{13}$, of course precludes labeling states by a parity quantum number, since $[H,\hat{P}_{23}]\ne 0$. In the symmetric case considered above, the QPT between LM and SC phases was characterized by a level crossing between parity-distinct states. We show below that the parity-broken model still supports LM and SC phases, but that the level crossing transition becomes instead a QPT of Kosterlitz-Thouless form.\cite{kt} 
The schematic phase diagram, Fig.~\ref{fig:pd}, still applies in the general case, but with the phase boundary now to be understood as a \emph{line} of SC critical end points, with the Kondo scale vanishing as the transition is approached from the SC phase.


\subsection{Effective models}\label{sec:asymmodel}

\begin{figure}
\begin{center}
\includegraphics[height=60mm]{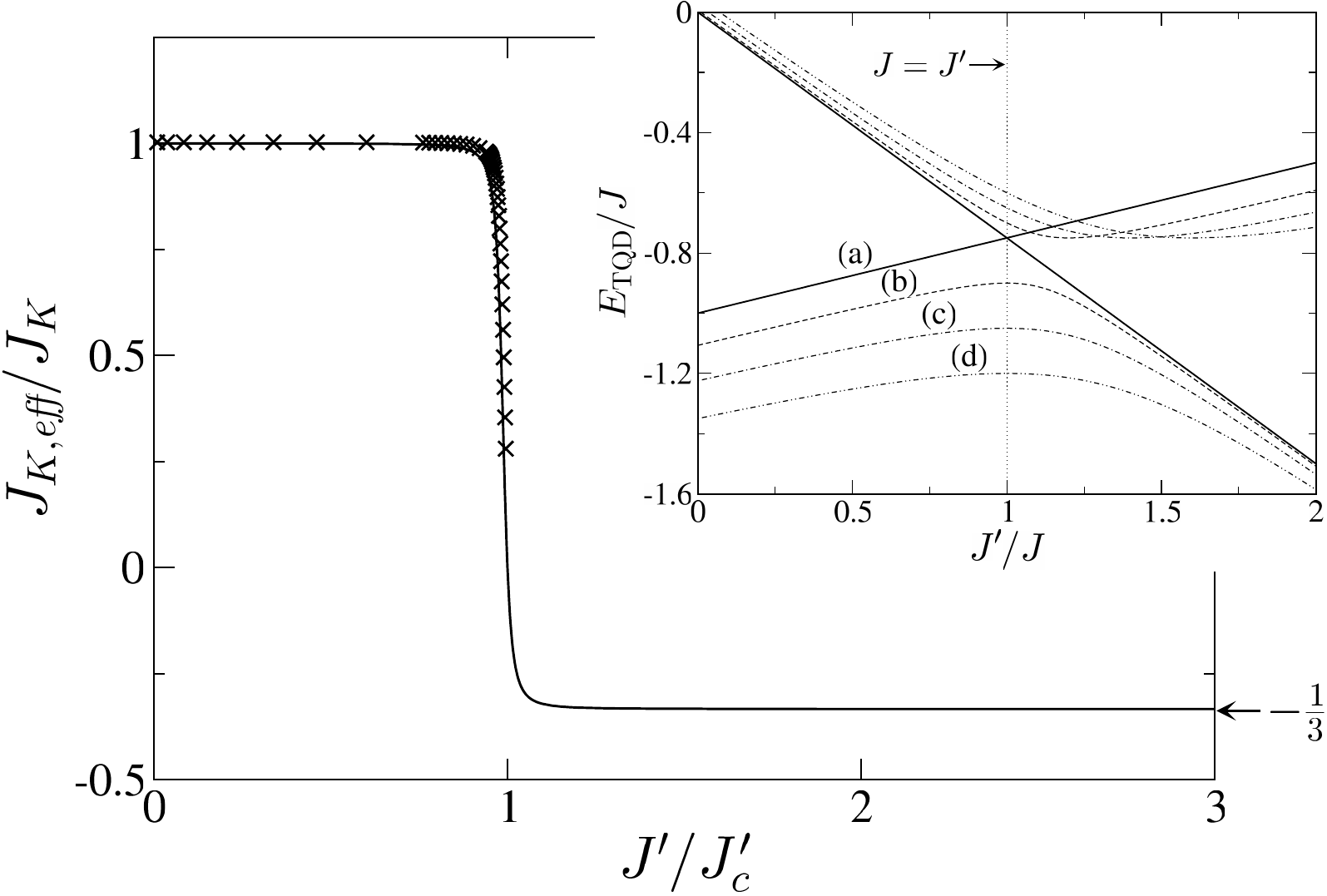}
\caption{\label{fig:largeJasym_model} 
Avoided crossing of the lowest TQD doublet states in the asymmetric
case shown in the inset, with $\delta/J=0$, $0.2$, $0.4$ and $0.6$ for
lines (a)--(d) respectively. Main panel shows the effective coupling of the lowest
doublet to the leads $J_{K,\text{eff}}/J_K$ vs $J'/J'_c$, for
$U/\pi\Gamma=10$, $t/\pi\Gamma=0.1$ and $\delta/\pi\Gamma=10^{-4}$,
obtained from Eq.~\ref{Jeffasym} (solid line); and extracted from NRG
data in the SC phase via $T_K\sim D\exp(-1/\rho J_{K,\text{eff}})$
(points).
}
\end{center}
\end{figure}

We focus here on the strongly correlated case of primary interest, $U \gg t_{ij}$, $\Gamma$. At temperatures $T\ll U$, each dot becomes essentially singly-occupied, and an effective TQD spin model analogous to Eq.~\ref{eq:spinmodel} can similarly be derived via a SWT upon perturbative elimination of virtual excitations to 2- and 4-electron TQD states. 
For ease of comparison with the parity-symmetric case, we retain $t_{13}\equiv t$ and $t_{23}\equiv t'$, defining the asymmetry in terms of  the parameter $x=t_{12}-t_{13}$. 
The resultant effective low-energy model then has the form
\begin{equation}\label{asymspin}
H_{\mathrm{spin}} = H_{\mathrm{leads}} +J \hat{\mathbf{S}}^{}_{1} \cdot (\hat{\mathbf{S}}^{}_{2}+\hat{\mathbf{S}}^{}_{3}) +J'\hat{\mathbf{S}}^{}_{2}\cdot \hat{\mathbf{S}}^{}_{3} +J^{}_{\mathrm{K}}\hat{\mathbf{S}}^{}_{1}\cdot \hat{\mathbf{s}}(0) + \delta \hat{\mathbf{S}}^{}_{1}\cdot \hat{\mathbf{S}}^{}_{2} \;
\end{equation}
where the effective exchange couplings $J$, $J'$ and $J_K$ are given in Eq.~\ref{eq:exchanges}; and where the asymmetry enters only through the final term, with 
\begin{equation}
\delta = 4\frac{(2tx+x^{2})}{U}.
\end{equation}
The low-energy manifold of the isolated TQD again comprises a pair of doublet states, denoted $|A;S^{z}\rangle$ and $|B;S^{z}\rangle$ (with an essentially irrelevant spin quartet occurring at higher energies). For any degree of asymmetry, these states may be expressed in terms of even- and odd-parity doublets $|\pm;S^z\rangle$ (themselves obtained in the symmetric limit $x=0=\delta$ considered above). They are given by
\begin{subequations}
\begin{align}
|A;S^z\rangle=& +\alpha |-;S^z\rangle + \beta |+;S^z\rangle \;,\\
|B;S^z\rangle=& -\beta |-;S^z\rangle + \alpha |+;S^z\rangle \;,
\end{align}
\end{subequations}
where 
\begin{equation}\label{alphabetadef}
\alpha^{2} = 1- \beta^2 = \frac{1}{2} \left ( 1+\frac{J-J'+\tfrac{1}{2} \delta}{\sqrt{(J-J'+\tfrac{1}{2} \delta)^{2}+\tfrac{3}{4}\delta^{2}}} \right ) \; .
\end{equation}
The energy difference between these isolated TQD doublet states is,
\begin{equation}\label{tqdE}
E_{\Delta} = E_{TQD}(B)-E_{TQD}(A) = \sqrt{(J-J'+\tfrac{1}{2} \delta)^{2}+\tfrac{3}{4}\delta^{2}}
\end{equation}
such that $E_{\Delta} > 0$ for all finite values of $J$, $J'$ and $\delta$ (or $t$, $t'$ and $x$). The effect of asymmetry is thus to turn the level crossing of TQD states at $\delta=0$ to an \emph{avoided} crossing when $\delta\ne 0$. There is in otherwords a single, unique doublet TQD ground state (the $|A;S^z\rangle$ state) for all model parameters in the asymmetric case. This is illustrated in the inset to Fig.~\ref{fig:largeJasym_model}, where we show the energies of the TQD doublet states as a function of $J'$ for different values of the asymmetry $\delta$ (the solid line (a) is for the symmetric case $\delta=0$).

The implication of Eq.~\ref{tqdE} is that below the temperature/energy scale $E_{\Delta}$, \emph{only} the TQD doublet state $|A;S^z\rangle$ is accessible. In consequence one can derive a low-energy effective model, valid for $T\ll E_{\Delta}$,  by projecting the spin model Eq.~\ref{asymspin} onto the ground state TQD doublet manifold spanned by $|A;S^z\rangle$. This can be done to first order in the Kondo coupling $J_K$, leading to
\begin{equation}\label{asymheff}
\begin{split}
H_{\textit{eff}} = & H_{\text{leads}}+ J_{\mathrm{K}} \sum_{S^z,{S^z}'} |A;S^z\rangle \langle A;S^z | \mathbf{\hat{S}}_{1}|A;{S^z}'\rangle \langle A;{S^z}' | \cdot \hat{\mathbf{s}}(0) \; ,\\
 = & H_{\text{leads}}+ J_{K,\textit{eff}} \mathbf{\hat{S}} \cdot \hat{\mathbf{s}}(0) \;,
\end{split}
\end{equation}
where $\mathbf{\hat{S}}$ is a spin-$\tfrac{1}{2}$ operator for the doublet TQD state $|A;S^z\rangle$; and where the effective exchange coupling to the leads follows as
\begin{equation}\label{Jeffasym}
J_{K,\textit{eff}} = J_{\mathrm{K}} \left( 1-\frac{4}{3}\alpha^{2} \right) \;.
\end{equation}
From Eq.~\ref{alphabetadef}, we immediately find that $J_{K,\textit{eff}}>0$ is antiferromagnetic for $J'>J$, but
is ferromagnetic, $J_{K,\textit{eff}}<0$, for $J'<J$ (and one also recovers asymptotically the results of the parity-symmetric case, $J_{K,\textit{eff}} \rightarrow -\tfrac{1}{3}J_K$, for $J\gg J', \delta$; and  
$J_{K,\textit{eff}} \rightarrow +J_K$ for $J'\gg J, \delta$).
The effective coupling as a function of $J'/J'_c$ is plotted in Fig.~\ref{fig:largeJasym_model} as the solid line; points are $J_{K,\textit{eff}}$ extracted from full NRG results, and show very good agreement.

Importantly, the antiferromagnetic effective Kondo coupling vanishes \emph{continuously} as  $(J'-J)\rightarrow 0^{+}$ [or $(t'-t)\rightarrow 0^{+}$] for any $|\delta|>0$,
\begin{equation}\label{vanish}
J_{K,\textit{eff}} = \frac{J_K}{2\delta} (J'-J) + \mathcal{O}(J'-J)^2 
\end{equation}
(in contrast to the discontinuous change in the effective Kondo coupling associated with a level crossing QPT, see e.g.~Fig.~\ref{fig:largeJsym}).
In consequence, in the full model we expect a Kondo SC phase for $t'>t'_c$ ($\simeq t$), with TQD degrees of freedom entirely quenched\cite{hewson} below a Kondo scale $T_K\sim D \exp(-1/\rho J_{K,\textit{eff}})$, which itself vanishes continuously as $(J'-J'_c)\rightarrow 0^{+}$ [or $(t'-t'_c)\rightarrow 0^{+}$)]:
\begin{equation}\label{ktscale}
\begin{split}
T_K \sim & D \exp \left [-\left (\frac{2\delta}{\rho J_K}\right ) \frac{1}{J'-J'_c} \right ]  \;, \\
\sim &   D \exp \left [-\left (\frac{\pi U x(2 t +x)}{8 \Gamma t}\right ) \frac{1}{t'-t'_c} \right ]  \;.
\end{split}
\end{equation}

The form of the vanishing Kondo scale as the transition is approached, Eq.~\ref{ktscale}, is that arising for a Kosterlitz-Thouless\cite{kt} (KT) transition. The point $t'=t'_c$ can thus be understood as the critical end point of a line of Kondo SC fixed points. For $t'<t'_c$, there is \emph{no} low-energy scale; and the residual TQD doublet degree of freedom remains free down to $T\rightarrow 0$ at this LM fixed point.


\begin{figure}[t]
\begin{center}
\includegraphics[height=60mm]{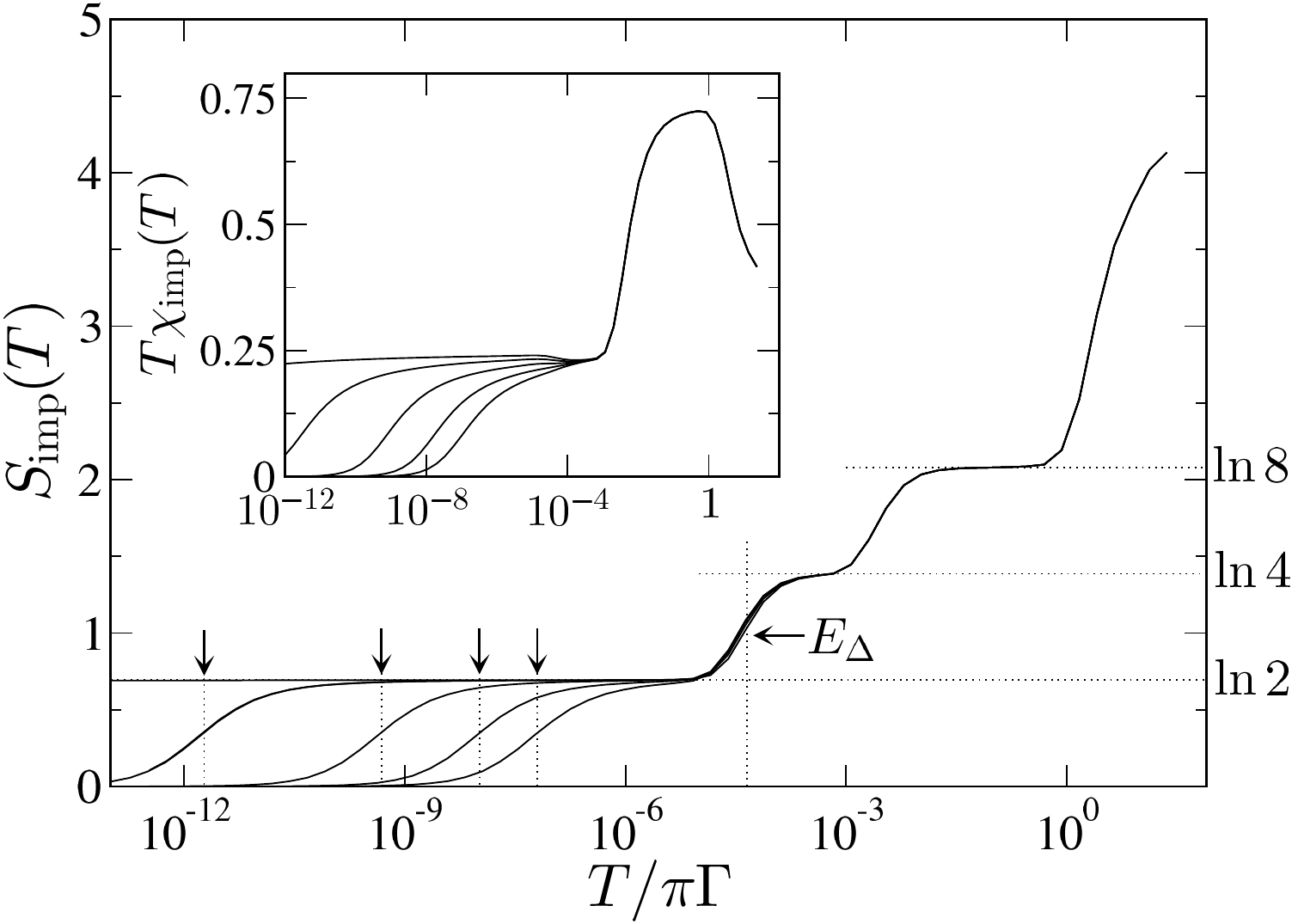}
\caption{\label{fig:largeJasym_td} Thermodynamics for the parity-broken model. TQD contribution to entropy $S_{\text{imp}}(T)$ (main panel) and magnetic susceptibility $T\chi_{\text{imp}}(T)$ (inset) vs $T/\pi \Gamma$. Plotted for $U/\pi\Gamma = 10$, $t/\pi\Gamma = 0.1$ and $\delta/\pi\Gamma = 10^{-4}$, varing $t'/\pi\Gamma=9.8\times 10^{-2}$, $9.78\times 10^{-2}$, $9.76\times 10^{-2}$, $9.74\times 10^{-2}$ and $9.72\times 10^{-2}$ to approach progressively the transition. Kondo scales $T_{K}$ are indicated by vertical arrows, and vanish continuously as the transition is approached. 
}
\end{center}
\end{figure}

\subsection{Kosterlitz-Thouless transition}\label{sec:asymKT}

\begin{figure}[t]
\begin{center}
\includegraphics[height=60mm]{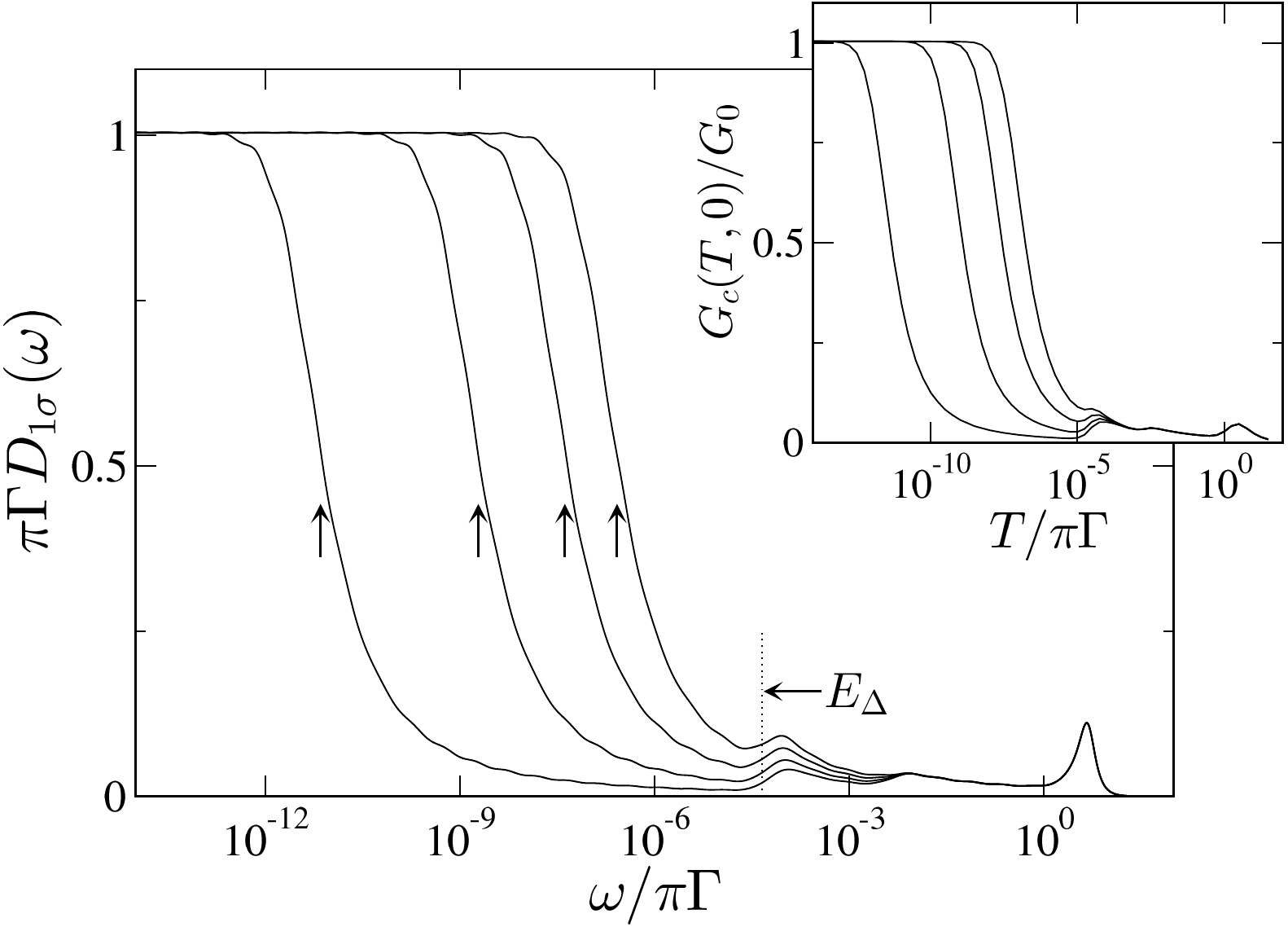}
\caption{\label{fig:largeJasym_spec} Single-particle spectrum of dot 1, $\pi\Gamma D_{1\sigma}(\omega)$ vs $\omega/\pi\Gamma$ (on a log-scale) at $T=0$, for systems with the same parameters as \ref{fig:largeJasym_td}. Vertical arrows indicate the Kondo scales $T_{K}$. Inset shows the temperature-dependence of the zero-bias conductance $G_c(T)/G_0$ for the same systems.
}
\end{center}
\end{figure}

This is explored further in Figs.~\ref{fig:largeJasym_td} and \ref{fig:largeJasym_spec}, containing NRG results for the full TQD model as the KT transition is approached from the SC phase, $t'>t'_c$.

Fig.~\ref{fig:largeJasym_td} shows thermodynamic quantities $S_{\text{imp}}$ and $T\chi_{\text{imp}}$ as a function of temperature. On the scale $T\sim U$, the dots become singly-occupied, and so the entropy drops to $S_{\text{imp}}=3\ln(2)$, with a corresponding tripled Curie law susceptibility, $T\chi_{\text{imp}}=\tfrac{3}{4}$. The quartet TQD state becomes inaccessible on the scale $T\sim J$ leaving the pair of quasi-degenerate TQD doublet states (and hence $S_{\mathrm{imp}} =\ln (4)$). 
The higher lying doublet is in turn projected out on the scale of $E_{\Delta}$. This is the LM fixed point, with $S_{\text{imp}}=\ln(2)$ and $T\chi_{\text{imp}}=\tfrac{1}{4}$. The lowest TQD doublet is then screened by the Kondo effect on the scale of $T_K$, which itself vanishes as the transition is approached (as evident directly from the figure). At the SC fixed point, TQD degrees of freedom are entirely quenched, giving $S_{\text{imp}}=0$ and $T\chi_{\text{imp}}=0$.

Similar behavior is observed in the $T=0$ spectral function of dot 1 shown in Fig.~\ref{fig:largeJasym_spec}, together (inset) with the $T$-dependence of the zero-bias conductance through the TQD.
On the energy scale $|\omega|\sim E_{\Delta}$, the higher energy TQD doublet state $|B;S^z\rangle$ is projected out, terminating the renormalization of its coupling to the leads. Instead, the lowest TQD $|A;S^z\rangle$ doublet flows to strong coupling, with the characteristic narrow Kondo resonance evident in the spectrum 
on the scale of $|\omega|\sim T_K$; and reaching the unitarity limit $\pi\Gamma D_{1\sigma}(\omega=0)=1$ that is characteristic\cite{hewson} of the 
strong coupling fixed point.
The Kondo resonance narrows progressively as $T_{K}$ diminishes and the QPT is approached from the SC phase; vanishing continuously `on the spot' at the transition itself, such that in (and throughout) the LM phase $\pi\Gamma D_{1\sigma}(\omega=0)=0$.
Corresponding behavior is naturally seen in the $T$-dependence of the conductance, which is determined by the local
spectrum as in Eq.~(\ref{mw}). Throughout the SC phase, it is likewise enhanced at low-temperatures $T\lesssim T_K$ due to the Kondo effect, with $G_c(T=0)/G_0=1$ in all cases. In the LM phase by contrast, where
$\pi\Gamma D_{1\sigma}(\omega=0)=0$, the $T=0$ conductance vanishes -- the conductance thus changing discontinuously as the transition is crossed, which appears to be a rather general signature of KT transitions in quantum dot and related systems.\cite{HofSch,garst,CCDD2005,CCDD2006,CW_multilevel}


\section{Summary and Discussion}\label{sec:concs}

A compact cluster of three quantum impurities, hybridizing apically with a single channel of host conduction electrons, has been shown to exhibit a rich range of physical behavior. Such a model may describe apex-coupled impurity trimers on metallic surfaces, or semiconductor TQD devices.

In the parity-symmetric case where two of the impurities/dots are in equivalent local environments, a level-crossing quantum phase transition separates a Kondo screened strong coupling (SC) phase and a free local moment (LM) phase. For strong interdot tunnel-coupling (and hence weak interactions), the transition in essence occurs between states with either 3 or 4 electrons occupying TQD molecular orbitals. For weaker couplings and strong dot electron correlations, sites of the TQD are essentially singly-occupied. Within this TQD spin-regime, the QPT between SC and LM phases again arises; although different mechanisms are uncovered, depending on the relative interdot tunneling strengths. 
 For weak interdot couplings in particular, the Kondo effect screens a single dot, with the residual TQD degrees of freedom forming either a local singlet (SC) or an underscreened spin-1 Kondo state (LM).
When parity symmetry is broken by distortions of the triangular TQD structure, we show that LM and SC phases are still supported --- with the transition between them now of Kosterlitz-Thouless type, such that the low-energy Kondo scale vanishes continuously as the transition is approached from the SC phase.

Given the ubiquity of SC and LM phases, an obvious question arises:
what characterizes -- and distinguishes -- such phases more generally? One answer is to note that the $T=0$ zero-bias conductance is pinned in the singly-occupied case to a unitarity value $G^{SC}_c(T=0)/G_0=1$ in the Kondo SC phase, but vanishes in the LM phase, $G^{LM}_c(T=0)/G_0=0$. The origin of this can be traced via Eq.~\ref{T0cond} to the scattering phase shift, $\delta$, experienced by conduction electrons in either phase. 

By a straightforward extension to the TQD model of the analysis given in Ref.\onlinecite{CW_multilevel}, a Friedel-Luttinger sum rule\cite{CW_multilevel,lutt_ward,lang} 
can be shown to relate exactly the phase shift to the excess charge ($n_{\text{imp}}(T=0)$) via
$\delta = \frac{\pi}{2}n_{\text{imp}}(T=0) + I_L $;
where the Luttinger integral $I_L$ is given by
\begin{equation}\label{lutt}
I_L = \text{Im} \text{Tr} \int^{0}_{-\infty} d\omega \frac{\partial \boldsymbol{\Sigma}_{\sigma}(\omega)}{\partial \omega}\mathbf{G}_{\sigma}(\omega) 
\end{equation}
(with $\mathbf{G}_{\sigma}(\omega)$ and $\boldsymbol{\Sigma}_{\sigma}(\omega)$ the $3\times 3$ matrices for the lead-coupled TQD Green functions and self-energies respectively). 
We find that the SC and LM phases are each characterized by a  distinct value for the `topological' quantity
$I_{L}$: from extensive NRG calculations of the Luttinger integral, and regardless of the bare underlying model parameters, we find $I_{L}=0$ to be characteristic of the SC phase (as indeed is well known for any Fermi liquid phase\cite{lutt_ward}), and $|I_{L}| =\frac{\pi}{2}$ to be equally 
characteristic of the LM phase. The latter is also precisely as found throughout the entire LM phase of a correlated 2-level quantum dot model\cite{CW_multilevel}, suggesting that the Luttinger integral is indeed a universal characteristic of a LM phase -- albeit that the fundamental reasons for this await an answer.


\acknowledgements
We are grateful for funding from EPSRC (UK) through grant EP/1032487/1.

One of us in particular (DEL) would also like to salute Peter Wolynes on the occasion of his 60$^{\mathrm{th}}$ birthday, and to acknowledge with gratitude the wealth of inspiration and insight provided by him over more than 30 years. Plus a lot of fun (which is not to be underestimated in science!).




\begin{thebibliography}{71}
\expandafter\ifx\csname natexlab\endcsname\relax\def\natexlab#1{#1}\fi
\expandafter\ifx\csname bibnamefont\endcsname\relax
  \def\bibnamefont#1{#1}\fi
\expandafter\ifx\csname bibfnamefont\endcsname\relax
  \def\bibfnamefont#1{#1}\fi
\expandafter\ifx\csname citenamefont\endcsname\relax
  \def\citenamefont#1{#1}\fi
\expandafter\ifx\csname url\endcsname\relax
  \def\url#1{\texttt{#1}}\fi
\expandafter\ifx\csname urlprefix\endcsname\relax\def\urlprefix{URL }\fi
\providecommand{\bibinfo}[2]{#2}
\providecommand{\eprint}[2][]{\url{#2}}

\bibitem[{\citenamefont{Hewson}(1993)}]{hewson}
\bibinfo{author}{\bibfnamefont{A.~C.} \bibnamefont{Hewson}},
  \emph{\bibinfo{title}{The {K}ondo Problem to Heavy Fermions}}
  (\bibinfo{publisher}{Cambridge University Press},
  \bibinfo{address}{Cambridge}, \bibinfo{year}{1993}).

\bibitem[{\citenamefont{Kondo}(1964)}]{kondo}
\bibinfo{author}{\bibfnamefont{J.}~\bibnamefont{Kondo}},
  \bibinfo{journal}{Prog. Theor. Phys.} \textbf{\bibinfo{volume}{32}},
  \bibinfo{pages}{37} (\bibinfo{year}{1964}).

\bibitem[{\citenamefont{Anderson}(1961)}]{siam}
\bibinfo{author}{\bibfnamefont{P.~W.} \bibnamefont{Anderson}},
  \bibinfo{journal}{Phys. Rev.} \textbf{\bibinfo{volume}{124}},
  \bibinfo{pages}{41} (\bibinfo{year}{1961}).

\bibitem[{\citenamefont{Wilson}(1975)}]{wilson}
\bibinfo{author}{\bibfnamefont{K.~G.} \bibnamefont{Wilson}},
  \bibinfo{journal}{Rev. Mod. Phys.} \textbf{\bibinfo{volume}{47}},
  \bibinfo{pages}{773} (\bibinfo{year}{1975}).

\bibitem[{\citenamefont{Mitchell
  et~al.}(2011{\natexlab{a}})\citenamefont{Mitchell, Becker, and
  Bulla}}]{akm:rginrs}
\bibinfo{author}{\bibfnamefont{A.~K.} \bibnamefont{Mitchell}},
  \bibinfo{author}{\bibfnamefont{M.}~\bibnamefont{Becker}}, \bibnamefont{and}
  \bibinfo{author}{\bibfnamefont{R.}~\bibnamefont{Bulla}},
  \bibinfo{journal}{Phys. Rev. B} \textbf{\bibinfo{volume}{84}},
  \bibinfo{pages}{115120} (\bibinfo{year}{2011}{\natexlab{a}}).

\bibitem[{\citenamefont{Krishnamurthy et~al.}(1980)\citenamefont{Krishnamurthy,
  Wilkins, and Wilson}}]{kww}
\bibinfo{author}{\bibfnamefont{H.~R.} \bibnamefont{Krishnamurthy}},
  \bibinfo{author}{\bibfnamefont{J.~W.} \bibnamefont{Wilkins}},
  \bibnamefont{and} \bibinfo{author}{\bibfnamefont{K.~G.}
  \bibnamefont{Wilson}}, \bibinfo{journal}{Phys. Rev. B}
  \textbf{\bibinfo{volume}{21}}, \bibinfo{pages}{1003, 1044}
  (\bibinfo{year}{1980}).

\bibitem[{\citenamefont{Jongeward and Wolynes}(1983)}]{wolynes}
\bibinfo{author}{\bibfnamefont{G.~A.} \bibnamefont{Jongeward}}
  \bibnamefont{and} \bibinfo{author}{\bibfnamefont{P.~G.}
  \bibnamefont{Wolynes}}, \bibinfo{journal}{J. Chem. Phys.}
  \textbf{\bibinfo{volume}{79}}, \bibinfo{pages}{3517} (\bibinfo{year}{1983}).

\bibitem[{\citenamefont{Withoff and Fradkin}(1990)}]{fradkin}
\bibinfo{author}{\bibfnamefont{D.}~\bibnamefont{Withoff}} \bibnamefont{and}
  \bibinfo{author}{\bibfnamefont{E.}~\bibnamefont{Fradkin}},
  \bibinfo{journal}{Phys. Rev. Lett.} \textbf{\bibinfo{volume}{64}},
  \bibinfo{pages}{1835} (\bibinfo{year}{1990}).

\bibitem[{\citenamefont{{Gonzalez-Buxton} and Ingersent}(1996)}]{sg_ingersent}
\bibinfo{author}{\bibfnamefont{C.}~\bibnamefont{{Gonzalez-Buxton}}}
  \bibnamefont{and}
  \bibinfo{author}{\bibfnamefont{K.}~\bibnamefont{Ingersent}},
  \bibinfo{journal}{Phys. Rev. B} \textbf{\bibinfo{volume}{54}},
  \bibinfo{pages}{15614} (\bibinfo{year}{1996}).

\bibitem[{\citenamefont{Logan and Glossop}(2000)}]{sg_lma}
\bibinfo{author}{\bibfnamefont{D.~E.} \bibnamefont{Logan}} \bibnamefont{and}
  \bibinfo{author}{\bibfnamefont{M.~T.} \bibnamefont{Glossop}},
  \bibinfo{journal}{J. Phys.: Condens. Matter} \textbf{\bibinfo{volume}{12}},
  \bibinfo{pages}{985} (\bibinfo{year}{2000}).

\bibitem[{\citenamefont{Bulla et~al.}(2000)\citenamefont{Bulla, Glossop, Logan,
  and Pruschke}}]{sg_bulla}
\bibinfo{author}{\bibfnamefont{R.}~\bibnamefont{Bulla}},
  \bibinfo{author}{\bibfnamefont{M.~T.} \bibnamefont{Glossop}},
  \bibinfo{author}{\bibfnamefont{D.~E.} \bibnamefont{Logan}}, \bibnamefont{and}
  \bibinfo{author}{\bibfnamefont{T.}~\bibnamefont{Pruschke}},
  \bibinfo{journal}{J. Phys.: Condens. Matter} \textbf{\bibinfo{volume}{12}},
  \bibinfo{pages}{4899} (\bibinfo{year}{2000}).

\bibitem[{\citenamefont{Glossop and Logan}(2003)}]{sg_lma2}
\bibinfo{author}{\bibfnamefont{M.~T.} \bibnamefont{Glossop}} \bibnamefont{and}
  \bibinfo{author}{\bibfnamefont{D.~E.} \bibnamefont{Logan}},
  \bibinfo{journal}{J. Phys.: Condens. Matter} \textbf{\bibinfo{volume}{15}},
  \bibinfo{pages}{7519} (\bibinfo{year}{2003}).

\bibitem[{\citenamefont{Fritz and Vojta}(2013)}]{rev:graphene}
\bibinfo{author}{\bibfnamefont{L.}~\bibnamefont{Fritz}} \bibnamefont{and}
  \bibinfo{author}{\bibfnamefont{M.}~\bibnamefont{Vojta}},
  \bibinfo{journal}{Rep. Prog. Phys.} \textbf{\bibinfo{volume}{76}},
  \bibinfo{pages}{032501} (\bibinfo{year}{2013}).

\bibitem[{\citenamefont{Mitchell and Fritz}(2013)}]{akm:graph}
\bibinfo{author}{\bibfnamefont{A.~K.} \bibnamefont{Mitchell}} \bibnamefont{and}
  \bibinfo{author}{\bibfnamefont{L.}~\bibnamefont{Fritz}},
  \bibinfo{journal}{Phys. Rev. B} \textbf{\bibinfo{volume}{88}},
  \bibinfo{pages}{075104} (\bibinfo{year}{2013}).

\bibitem[{\citenamefont{Fritz and Vojta}(2005)}]{dwavesc}
\bibinfo{author}{\bibfnamefont{L.}~\bibnamefont{Fritz}} \bibnamefont{and}
  \bibinfo{author}{\bibfnamefont{M.}~\bibnamefont{Vojta}},
  \bibinfo{journal}{Phys. Rev. B} \textbf{\bibinfo{volume}{72}},
  \bibinfo{pages}{212510} (\bibinfo{year}{2005}).

\bibitem[{\citenamefont{Mitchell et~al.}(2013)\citenamefont{Mitchell,
  Schurucht, Vojta, and Fritz}}]{akm:qpiti}
\bibinfo{author}{\bibfnamefont{A.~K.} \bibnamefont{Mitchell}},
  \bibinfo{author}{\bibfnamefont{D.}~\bibnamefont{Schurucht}},
  \bibinfo{author}{\bibfnamefont{M.}~\bibnamefont{Vojta}}, \bibnamefont{and}
  \bibinfo{author}{\bibfnamefont{L.}~\bibnamefont{Fritz}},
  \bibinfo{journal}{Phys. Rev. B} \textbf{\bibinfo{volume}{87}},
  \bibinfo{pages}{075430} (\bibinfo{year}{2013}).

\bibitem[{\citenamefont{Zitko}(2010)}]{zitko_ti}
\bibinfo{author}{\bibfnamefont{R.}~\bibnamefont{Zitko}},
  \bibinfo{journal}{Phys. Rev. B} \textbf{\bibinfo{volume}{81}},
  \bibinfo{pages}{241414(R)} (\bibinfo{year}{2010}).

\bibitem[{jon()}]{jones}
\bibinfo{note}{B.~A.~Jones, C.~M. Varma, and J. W. Wilkins, Phys. Rev. Lett.
  \textbf{61}, 125 (1988); B. A. Jones, Physica B (Amsterdam) \textbf{171}, 53
  (1991).}

\bibitem[{CFT()}]{CFT2IKM}
\bibinfo{note}{I.~Affleck and A.~W.~W.~Ludwig, Phys. Rev. Lett. \textbf{68},
  1046 (1992); I.~Affleck, A.~W.~W.~Ludwig and B.~A.~Jones, Phys. Rev. B
  \textbf{52}, 9528 (1995).}

\bibitem[{\citenamefont{Jayatilaka et~al.}(2011)\citenamefont{Jayatilaka,
  Galpin, and Logan}}]{DEL_2iam}
\bibinfo{author}{\bibfnamefont{F.~W.} \bibnamefont{Jayatilaka}},
  \bibinfo{author}{\bibfnamefont{M.~R.} \bibnamefont{Galpin}},
  \bibnamefont{and} \bibinfo{author}{\bibfnamefont{D.~E.} \bibnamefont{Logan}},
  \bibinfo{journal}{Phys. Rev. B} \textbf{\bibinfo{volume}{84}},
  \bibinfo{pages}{115111} (\bibinfo{year}{2011}).

\bibitem[{\citenamefont{Sela et~al.}(2011)\citenamefont{Sela, Mitchell, and
  Fritz}}]{akm:exactNFL}
\bibinfo{author}{\bibfnamefont{E.}~\bibnamefont{Sela}},
  \bibinfo{author}{\bibfnamefont{A.~K.} \bibnamefont{Mitchell}},
  \bibnamefont{and} \bibinfo{author}{\bibfnamefont{L.}~\bibnamefont{Fritz}},
  \bibinfo{journal}{Phys. Rev. Lett.} \textbf{\bibinfo{volume}{106}},
  \bibinfo{pages}{147202} (\bibinfo{year}{2011}).

\bibitem[{\citenamefont{Mitchell and Sela}(2012)}]{akm:finiteT}
\bibinfo{author}{\bibfnamefont{A.~K.} \bibnamefont{Mitchell}} \bibnamefont{and}
  \bibinfo{author}{\bibfnamefont{E.}~\bibnamefont{Sela}},
  \bibinfo{journal}{Phys. Rev. B} \textbf{\bibinfo{volume}{85}},
  \bibinfo{pages}{235127} (\bibinfo{year}{2012}).

\bibitem[{\citenamefont{Mitchell et~al.}(2012)\citenamefont{Mitchell, Sela, and
  Logan}}]{akm:2ckin2ik}
\bibinfo{author}{\bibfnamefont{A.~K.} \bibnamefont{Mitchell}},
  \bibinfo{author}{\bibfnamefont{E.}~\bibnamefont{Sela}}, \bibnamefont{and}
  \bibinfo{author}{\bibfnamefont{D.~E.} \bibnamefont{Logan}},
  \bibinfo{journal}{Phys. Rev. Lett.} \textbf{\bibinfo{volume}{108}},
  \bibinfo{pages}{086405} (\bibinfo{year}{2012}).

\bibitem[{\citenamefont{Jayaprakash et~al.}(1981)\citenamefont{Jayaprakash,
  Krishnamurthy, and Wilkins}}]{2imp_krish}
\bibinfo{author}{\bibfnamefont{C.}~\bibnamefont{Jayaprakash}},
  \bibinfo{author}{\bibfnamefont{H.~R.} \bibnamefont{Krishnamurthy}},
  \bibnamefont{and} \bibinfo{author}{\bibfnamefont{J.~W.}
  \bibnamefont{Wilkins}}, \bibinfo{journal}{Phys. Rev. Lett.}
  \textbf{\bibinfo{volume}{47}}, \bibinfo{pages}{737} (\bibinfo{year}{1981}).

\bibitem[{\citenamefont{van~der Wiel et~al.}(2003)}]{dd_rev}
\bibinfo{author}{\bibfnamefont{W.~G.} \bibnamefont{van~der Wiel}}
  \bibnamefont{et~al.}, \bibinfo{journal}{Rev. Mod. Phys.}
  \textbf{\bibinfo{volume}{75}}, \bibinfo{pages}{1} (\bibinfo{year}{2003}).

\bibitem[{\citenamefont{Chang and Chen}(2009)}]{Chang_rev}
\bibinfo{author}{\bibfnamefont{A.~M.} \bibnamefont{Chang}} \bibnamefont{and}
  \bibinfo{author}{\bibfnamefont{J.~C.} \bibnamefont{Chen}},
  \bibinfo{journal}{Rep. Prog. Phys.} \textbf{\bibinfo{volume}{72}},
  \bibinfo{pages}{096501} (\bibinfo{year}{2009}).

\bibitem[{\citenamefont{Merino et~al.}(2009)\citenamefont{Merino, Borda, and
  Simon}}]{simondimer}
\bibinfo{author}{\bibfnamefont{J.}~\bibnamefont{Merino}},
  \bibinfo{author}{\bibfnamefont{L.}~\bibnamefont{Borda}}, \bibnamefont{and}
  \bibinfo{author}{\bibfnamefont{P.}~\bibnamefont{Simon}},
  \bibinfo{journal}{Europhys. Lett.} \textbf{\bibinfo{volume}{85}},
  \bibinfo{pages}{47002} (\bibinfo{year}{2009}).

\bibitem[{\citenamefont{Chorley et~al.}(2012)\citenamefont{Chorley, Galpin,
  Jayatilaka, Smith, Logan, and Buitelaar}}]{del_cntdqd}
\bibinfo{author}{\bibfnamefont{S.~J.} \bibnamefont{Chorley}},
  \bibinfo{author}{\bibfnamefont{M.~R.} \bibnamefont{Galpin}},
  \bibinfo{author}{\bibfnamefont{F.~W.} \bibnamefont{Jayatilaka}},
  \bibinfo{author}{\bibfnamefont{C.~G.} \bibnamefont{Smith}},
  \bibinfo{author}{\bibfnamefont{D.~E.} \bibnamefont{Logan}}, \bibnamefont{and}
  \bibinfo{author}{\bibfnamefont{M.~R.} \bibnamefont{Buitelaar}},
  \bibinfo{journal}{Phys. Rev. Lett.} \textbf{\bibinfo{volume}{109}},
  \bibinfo{pages}{156804} (\bibinfo{year}{2012}).

\bibitem[{\citenamefont{Mitchell et~al.}(2006)\citenamefont{Mitchell, Galpin,
  and Logan}}]{akm:ccdqd}
\bibinfo{author}{\bibfnamefont{A.~K.} \bibnamefont{Mitchell}},
  \bibinfo{author}{\bibfnamefont{M.~R.} \bibnamefont{Galpin}},
  \bibnamefont{and} \bibinfo{author}{\bibfnamefont{D.~E.} \bibnamefont{Logan}},
  \bibinfo{journal}{Europhys. Lett.} \textbf{\bibinfo{volume}{76}},
  \bibinfo{pages}{95} (\bibinfo{year}{2006}).

\bibitem[{\citenamefont{Jamneala et~al.}(2001)\citenamefont{Jamneala, Madhavan,
  and Crommie}}]{cr3:expt}
\bibinfo{author}{\bibfnamefont{T.}~\bibnamefont{Jamneala}},
  \bibinfo{author}{\bibfnamefont{V.}~\bibnamefont{Madhavan}}, \bibnamefont{and}
  \bibinfo{author}{\bibfnamefont{M.~F.} \bibnamefont{Crommie}},
  \bibinfo{journal}{Phys. Rev. Lett.} \textbf{\bibinfo{volume}{87}},
  \bibinfo{pages}{256804} (\bibinfo{year}{2001}).

\bibitem[{\citenamefont{Ingersent et~al.}(2005)\citenamefont{Ingersent, Ludwig,
  and Affleck}}]{cr3:theory}
\bibinfo{author}{\bibfnamefont{K.}~\bibnamefont{Ingersent}},
  \bibinfo{author}{\bibfnamefont{A.~W.~W.} \bibnamefont{Ludwig}},
  \bibnamefont{and} \bibinfo{author}{\bibfnamefont{I.}~\bibnamefont{Affleck}},
  \bibinfo{journal}{Phys. Rev. Lett.} \textbf{\bibinfo{volume}{95}},
  \bibinfo{pages}{257204} (\bibinfo{year}{2005}).

\bibitem[{\citenamefont{Mitchell et~al.}(2009)\citenamefont{Mitchell, Jarrold,
  and Logan}}]{akm:tqd1ch}
\bibinfo{author}{\bibfnamefont{A.~K.} \bibnamefont{Mitchell}},
  \bibinfo{author}{\bibfnamefont{T.~F.} \bibnamefont{Jarrold}},
  \bibnamefont{and} \bibinfo{author}{\bibfnamefont{D.~E.} \bibnamefont{Logan}},
  \bibinfo{journal}{Phys. Rev. B} \textbf{\bibinfo{volume}{79}},
  \bibinfo{pages}{085124} (\bibinfo{year}{2009}).

\bibitem[{\citenamefont{Mitchell and Logan}(2010)}]{akm:tqd2ch}
\bibinfo{author}{\bibfnamefont{A.~K.} \bibnamefont{Mitchell}} \bibnamefont{and}
  \bibinfo{author}{\bibfnamefont{D.~E.} \bibnamefont{Logan}},
  \bibinfo{journal}{Phys. Rev. B} \textbf{\bibinfo{volume}{81}},
  \bibinfo{pages}{075126} (\bibinfo{year}{2010}).

\bibitem[{\citenamefont{Uchihashi et~al.}(2008)\citenamefont{Uchihashi, Zhang,
  Kr\"{o}ger, and Berndt}}]{Cotrimer}
\bibinfo{author}{\bibfnamefont{T.}~\bibnamefont{Uchihashi}},
  \bibinfo{author}{\bibfnamefont{J.}~\bibnamefont{Zhang}},
  \bibinfo{author}{\bibfnamefont{J.}~\bibnamefont{Kr\"{o}ger}},
  \bibnamefont{and} \bibinfo{author}{\bibfnamefont{R.}~\bibnamefont{Berndt}},
  \bibinfo{journal}{Phys. Rev. B} \textbf{\bibinfo{volume}{78}},
  \bibinfo{pages}{033402} (\bibinfo{year}{2008}).

\bibitem[{\citenamefont{Lazarovits et~al.}(2005)\citenamefont{Lazarovits,
  Simon, Zar\'{a}nd, and Szunyogh}}]{lazarovitsadatom}
\bibinfo{author}{\bibfnamefont{B.}~\bibnamefont{Lazarovits}},
  \bibinfo{author}{\bibfnamefont{P.}~\bibnamefont{Simon}},
  \bibinfo{author}{\bibfnamefont{G.}~\bibnamefont{Zar\'{a}nd}},
  \bibnamefont{and} \bibinfo{author}{\bibfnamefont{L.}~\bibnamefont{Szunyogh}},
  \bibinfo{journal}{Phys. Rev. Lett.} \textbf{\bibinfo{volume}{95}},
  \bibinfo{eid}{077202} (\bibinfo{year}{2005}).

\bibitem[{\citenamefont{Wang}(2007)}]{wangtqd}
\bibinfo{author}{\bibfnamefont{W.~Z.} \bibnamefont{Wang}},
  \bibinfo{journal}{Phys. Rev. B} \textbf{\bibinfo{volume}{76}},
  \bibinfo{eid}{115114} (\bibinfo{year}{2007}).

\bibitem[{\citenamefont{Xiong et~al.}(2012)\citenamefont{Xiong, Huang, and
  Wang}}]{wangtqd2}
\bibinfo{author}{\bibfnamefont{Y.~C.} \bibnamefont{Xiong}},
  \bibinfo{author}{\bibfnamefont{J.}~\bibnamefont{Huang}}, \bibnamefont{and}
  \bibinfo{author}{\bibfnamefont{W.-Z.} \bibnamefont{Wang}},
  \bibinfo{journal}{J. Phys.: Condens. Matter} \textbf{\bibinfo{volume}{24}},
  \bibinfo{pages}{455604} (\bibinfo{year}{2012}).

\bibitem[{\citenamefont{\v{Z}itko and Bon\v{c}a}(2008)}]{zitkoTQD2ch}
\bibinfo{author}{\bibfnamefont{R.}~\bibnamefont{\v{Z}itko}} \bibnamefont{and}
  \bibinfo{author}{\bibfnamefont{J.}~\bibnamefont{Bon\v{c}a}},
  \bibinfo{journal}{Phys. Rev. B} \textbf{\bibinfo{volume}{77}},
  \bibinfo{pages}{245112} (\bibinfo{year}{2008}).

\bibitem[{\citenamefont{Oguri et~al.}(2011)\citenamefont{Oguri, Amaha,
  Nishikawa, Numata, Shimamoto, Hewson, and Tarucha}}]{hewsontqd1}
\bibinfo{author}{\bibfnamefont{A.}~\bibnamefont{Oguri}},
  \bibinfo{author}{\bibfnamefont{S.}~\bibnamefont{Amaha}},
  \bibinfo{author}{\bibfnamefont{Y.}~\bibnamefont{Nishikawa}},
  \bibinfo{author}{\bibfnamefont{T.}~\bibnamefont{Numata}},
  \bibinfo{author}{\bibfnamefont{M.}~\bibnamefont{Shimamoto}},
  \bibinfo{author}{\bibfnamefont{A.~C.} \bibnamefont{Hewson}},
  \bibnamefont{and} \bibinfo{author}{\bibfnamefont{S.}~\bibnamefont{Tarucha}},
  \bibinfo{journal}{Phys. Rev. B} \textbf{\bibinfo{volume}{83}},
  \bibinfo{pages}{205304} (\bibinfo{year}{2011}).

\bibitem[{\citenamefont{Numata et~al.}(2009)\citenamefont{Numata, Nisikawa,
  Oguri, and Hewson}}]{hewsontqd2}
\bibinfo{author}{\bibfnamefont{T.}~\bibnamefont{Numata}},
  \bibinfo{author}{\bibfnamefont{T.}~\bibnamefont{Nisikawa}},
  \bibinfo{author}{\bibfnamefont{A.}~\bibnamefont{Oguri}}, \bibnamefont{and}
  \bibinfo{author}{\bibfnamefont{A.~C.} \bibnamefont{Hewson}},
  \bibinfo{journal}{Phys. Rev. B} \textbf{\bibinfo{volume}{80}},
  \bibinfo{pages}{155330} (\bibinfo{year}{2009}).

\bibitem[{\citenamefont{Seo et~al.}(2013)}]{tqdexpt}
\bibinfo{author}{\bibfnamefont{M.}~\bibnamefont{Seo}} \bibnamefont{et~al.},
  \bibinfo{journal}{Phys. Rev. Lett.} \textbf{\bibinfo{volume}{110}},
  \bibinfo{pages}{046803} (\bibinfo{year}{2013}).

\bibitem[{\citenamefont{Kudasov and Uzdin}(2002)}]{cr3:uzdin}
\bibinfo{author}{\bibfnamefont{Y.~B.} \bibnamefont{Kudasov}} \bibnamefont{and}
  \bibinfo{author}{\bibfnamefont{V.~M.} \bibnamefont{Uzdin}},
  \bibinfo{journal}{Phys. Rev. Lett.} \textbf{\bibinfo{volume}{89}},
  \bibinfo{pages}{276802} (\bibinfo{year}{2002}).

\bibitem[{\citenamefont{Gotsis et~al.}(2006)\citenamefont{Gotsis, Kioussis, and
  Papaconstantopoulos}}]{cr3:Gotsis}
\bibinfo{author}{\bibfnamefont{H.~J.} \bibnamefont{Gotsis}},
  \bibinfo{author}{\bibfnamefont{N.}~\bibnamefont{Kioussis}}, \bibnamefont{and}
  \bibinfo{author}{\bibfnamefont{D.~A.} \bibnamefont{Papaconstantopoulos}},
  \bibinfo{journal}{Phys. Rev. B} \textbf{\bibinfo{volume}{73}},
  \bibinfo{pages}{014436} (\bibinfo{year}{2006}).

\bibitem[{\citenamefont{Antal et~al.}(2008)\citenamefont{Antal, Lazarovits,
  Szunyogh, Ujfalussy, and Weinberger}}]{cr3:frust}
\bibinfo{author}{\bibfnamefont{A.}~\bibnamefont{Antal}},
  \bibinfo{author}{\bibfnamefont{B.}~\bibnamefont{Lazarovits}},
  \bibinfo{author}{\bibfnamefont{L.}~\bibnamefont{Szunyogh}},
  \bibinfo{author}{\bibfnamefont{B.}~\bibnamefont{Ujfalussy}},
  \bibnamefont{and}
  \bibinfo{author}{\bibfnamefont{P.}~\bibnamefont{Weinberger}},
  \bibinfo{journal}{Phys. Rev. B} \textbf{\bibinfo{volume}{77}},
  \bibinfo{pages}{174429} (\bibinfo{year}{2008}).

\bibitem[{\citenamefont{Schr\"{o}er et~al.}(2007)\citenamefont{Schr\"{o}er,
  Greentree, Gaudreau, Eberl, Hollenberg, Kotthaus, and Ludwig}}]{schroer}
\bibinfo{author}{\bibfnamefont{D.}~\bibnamefont{Schr\"{o}er}},
  \bibinfo{author}{\bibfnamefont{A.~D.} \bibnamefont{Greentree}},
  \bibinfo{author}{\bibfnamefont{L.}~\bibnamefont{Gaudreau}},
  \bibinfo{author}{\bibfnamefont{K.}~\bibnamefont{Eberl}},
  \bibinfo{author}{\bibfnamefont{L.~C.~L.} \bibnamefont{Hollenberg}},
  \bibinfo{author}{\bibfnamefont{J.~P.} \bibnamefont{Kotthaus}},
  \bibnamefont{and} \bibinfo{author}{\bibfnamefont{S.}~\bibnamefont{Ludwig}},
  \bibinfo{journal}{Phys. Rev. B} \textbf{\bibinfo{volume}{76}},
  \bibinfo{pages}{075306} (\bibinfo{year}{2007}).

\bibitem[{\citenamefont{Vidan et~al.}(2005)\citenamefont{Vidan, Westervelt,
  Stopa, Hanson, and Gossard}}]{vidan3dot}
\bibinfo{author}{\bibfnamefont{A.}~\bibnamefont{Vidan}},
  \bibinfo{author}{\bibfnamefont{R.}~\bibnamefont{Westervelt}},
  \bibinfo{author}{\bibfnamefont{M.}~\bibnamefont{Stopa}},
  \bibinfo{author}{\bibfnamefont{M.}~\bibnamefont{Hanson}}, \bibnamefont{and}
  \bibinfo{author}{\bibfnamefont{A.}~\bibnamefont{Gossard}},
  \bibinfo{journal}{J. Supercond. Incorp. Novel Magn.}
  \textbf{\bibinfo{volume}{18}}, \bibinfo{pages}{223} (\bibinfo{year}{2005}).

\bibitem[{\citenamefont{Rogge and Haug}(2008)}]{rogge3dot}
\bibinfo{author}{\bibfnamefont{M.~C.} \bibnamefont{Rogge}} \bibnamefont{and}
  \bibinfo{author}{\bibfnamefont{R.~J.} \bibnamefont{Haug}},
  \bibinfo{journal}{Phys. Rev. B} \textbf{\bibinfo{volume}{77}},
  \bibinfo{pages}{193306} (\bibinfo{year}{2008}).

\bibitem[{\citenamefont{Gaudreau et~al.}(2009)\citenamefont{Gaudreau, Kam,
  Granger, Studenikin, Zawadzki, and Sachrajda}}]{Gaudreau}
\bibinfo{author}{\bibfnamefont{L.}~\bibnamefont{Gaudreau}},
  \bibinfo{author}{\bibfnamefont{A.}~\bibnamefont{Kam}},
  \bibinfo{author}{\bibfnamefont{G.}~\bibnamefont{Granger}},
  \bibinfo{author}{\bibfnamefont{S.~A.} \bibnamefont{Studenikin}},
  \bibinfo{author}{\bibfnamefont{P.}~\bibnamefont{Zawadzki}}, \bibnamefont{and}
  \bibinfo{author}{\bibfnamefont{A.~S.} \bibnamefont{Sachrajda}},
  \bibinfo{journal}{Appl. Phys. Lett.} \textbf{\bibinfo{volume}{95}},
  \bibinfo{pages}{193101} (\bibinfo{year}{2009}).

\bibitem[{qd:()}]{qd:expt}
\bibinfo{note}{D. Goldhaber-Gordon, H. Shtrikman, D. Mahalu, D. Abusch-Magder,
  U. Meirav, and M. A. Kastner, Nature \textbf{391}, 156 (1998); S. M.
  Cronenwett, T. H. Oosterkamp, and L. P. Kouwenhoven, Science \textbf{540},
  281 (1998).}

\bibitem[{\citenamefont{Kastner}(1993)}]{artatom}
\bibinfo{author}{\bibfnamefont{M.~A.} \bibnamefont{Kastner}},
  \bibinfo{journal}{Physics Today} \textbf{\bibinfo{volume}{46}},
  \bibinfo{pages}{24} (\bibinfo{year}{1993}).

\bibitem[{\citenamefont{Jeong et~al.}(2001)\citenamefont{Jeong, Chang, and
  Melloch}}]{artmol}
\bibinfo{author}{\bibfnamefont{H.}~\bibnamefont{Jeong}},
  \bibinfo{author}{\bibfnamefont{A.~M.} \bibnamefont{Chang}}, \bibnamefont{and}
  \bibinfo{author}{\bibfnamefont{M.~R.} \bibnamefont{Melloch}},
  \bibinfo{journal}{Science} \textbf{\bibinfo{volume}{293}},
  \bibinfo{pages}{2221} (\bibinfo{year}{2001}).

\bibitem[{\citenamefont{Blick et~al.}(1996)\citenamefont{Blick, Haug, Weis,
  Pfannkuche, Klitzing, and Eberl}}]{blick}
\bibinfo{author}{\bibfnamefont{R.~H.} \bibnamefont{Blick}},
  \bibinfo{author}{\bibfnamefont{R.~J.} \bibnamefont{Haug}},
  \bibinfo{author}{\bibfnamefont{J.}~\bibnamefont{Weis}},
  \bibinfo{author}{\bibfnamefont{D.}~\bibnamefont{Pfannkuche}},
  \bibinfo{author}{\bibfnamefont{K.~v.} \bibnamefont{Klitzing}},
  \bibnamefont{and} \bibinfo{author}{\bibfnamefont{K.}~\bibnamefont{Eberl}},
  \bibinfo{journal}{Phys. Rev. B} \textbf{\bibinfo{volume}{53}},
  \bibinfo{pages}{7899} (\bibinfo{year}{1996}).

\bibitem[{\citenamefont{Kosterlitz and Thouless}(1973)}]{kt}
\bibinfo{author}{\bibfnamefont{J.~M.} \bibnamefont{Kosterlitz}}
  \bibnamefont{and} \bibinfo{author}{\bibfnamefont{D.~J.}
  \bibnamefont{Thouless}}, \bibinfo{journal}{J. Phys. C}
  \textbf{\bibinfo{volume}{6}}, \bibinfo{pages}{1181} (\bibinfo{year}{1973}).

\bibitem[{\citenamefont{Bulla et~al.}(2008)\citenamefont{Bulla, Costi, and
  Pruschke}}]{nrg:rev}
\bibinfo{author}{\bibfnamefont{R.}~\bibnamefont{Bulla}},
  \bibinfo{author}{\bibfnamefont{T.}~\bibnamefont{Costi}}, \bibnamefont{and}
  \bibinfo{author}{\bibfnamefont{T.}~\bibnamefont{Pruschke}},
  \bibinfo{journal}{Rev. Mod. Phys.} \textbf{\bibinfo{volume}{80}},
  \bibinfo{pages}{395} (\bibinfo{year}{2008}).

\bibitem[{\citenamefont{Peters et~al.}(2006)\citenamefont{Peters, Pruschke, and
  Anders}}]{asbasis}
\bibinfo{author}{\bibfnamefont{R.}~\bibnamefont{Peters}},
  \bibinfo{author}{\bibfnamefont{T.}~\bibnamefont{Pruschke}}, \bibnamefont{and}
  \bibinfo{author}{\bibfnamefont{F.~B.} \bibnamefont{Anders}},
  \bibinfo{journal}{Phys. Rev. B} \textbf{\bibinfo{volume}{74}},
  \bibinfo{eid}{245114} (\bibinfo{year}{2006}).

\bibitem[{\citenamefont{Weichselbaum and von Delft}(2007)}]{fdmnrg}
\bibinfo{author}{\bibfnamefont{A.}~\bibnamefont{Weichselbaum}}
  \bibnamefont{and} \bibinfo{author}{\bibfnamefont{J.}~\bibnamefont{von
  Delft}}, \bibinfo{journal}{Phys. Rev. Lett.} \textbf{\bibinfo{volume}{99}},
  \bibinfo{eid}{076402} (\bibinfo{year}{2007}).

\bibitem[{\citenamefont{Anders and Schiller}(2005)}]{asbasisprl}
\bibinfo{author}{\bibfnamefont{F.~B.} \bibnamefont{Anders}} \bibnamefont{and}
  \bibinfo{author}{\bibfnamefont{A.}~\bibnamefont{Schiller}},
  \bibinfo{journal}{Phys. Rev. Lett.} \textbf{\bibinfo{volume}{95}},
  \bibinfo{pages}{196801} (\bibinfo{year}{2005}).

\bibitem[{\citenamefont{Meir and Wingreen}(1992)}]{meir}
\bibinfo{author}{\bibfnamefont{Y.}~\bibnamefont{Meir}} \bibnamefont{and}
  \bibinfo{author}{\bibfnamefont{N.~S.} \bibnamefont{Wingreen}},
  \bibinfo{journal}{Phys. Rev. Lett.} \textbf{\bibinfo{volume}{68}},
  \bibinfo{pages}{2512} (\bibinfo{year}{1992}).

\bibitem[{\citenamefont{Schrieffer and Wolff}(1966)}]{sw}
\bibinfo{author}{\bibfnamefont{J.~R.} \bibnamefont{Schrieffer}}
  \bibnamefont{and} \bibinfo{author}{\bibfnamefont{P.~A.} \bibnamefont{Wolff}},
  \bibinfo{journal}{Phys. Rev.} \textbf{\bibinfo{volume}{149}},
  \bibinfo{pages}{491} (\bibinfo{year}{1966}).

\bibitem[{\citenamefont{Mehta et~al.}(2005)\citenamefont{Mehta, Andrei,
  Coleman, Borda, and Zarand}}]{lmcor1}
\bibinfo{author}{\bibfnamefont{P.}~\bibnamefont{Mehta}},
  \bibinfo{author}{\bibfnamefont{N.}~\bibnamefont{Andrei}},
  \bibinfo{author}{\bibfnamefont{P.}~\bibnamefont{Coleman}},
  \bibinfo{author}{\bibfnamefont{L.}~\bibnamefont{Borda}}, \bibnamefont{and}
  \bibinfo{author}{\bibfnamefont{G.}~\bibnamefont{Zarand}},
  \bibinfo{journal}{Phys. Rev. B} \textbf{\bibinfo{volume}{72}},
  \bibinfo{pages}{014430} (\bibinfo{year}{2005}).

\bibitem[{\citenamefont{Koller et~al.}(2005)\citenamefont{Koller, Hewson, and
  Meyer}}]{lmcor2}
\bibinfo{author}{\bibfnamefont{W.}~\bibnamefont{Koller}},
  \bibinfo{author}{\bibfnamefont{A.~C.} \bibnamefont{Hewson}},
  \bibnamefont{and} \bibinfo{author}{\bibfnamefont{D.}~\bibnamefont{Meyer}},
  \bibinfo{journal}{Phys. Rev. B} \textbf{\bibinfo{volume}{72}},
  \bibinfo{pages}{045117} (\bibinfo{year}{2005}).

\bibitem[{\citenamefont{Mitchell
  et~al.}(2011{\natexlab{b}})\citenamefont{Mitchell, Logan, and
  Krishnamurthy}}]{akm:oddimp}
\bibinfo{author}{\bibfnamefont{A.~K.} \bibnamefont{Mitchell}},
  \bibinfo{author}{\bibfnamefont{D.~E.} \bibnamefont{Logan}}, \bibnamefont{and}
  \bibinfo{author}{\bibfnamefont{H.~R.} \bibnamefont{Krishnamurthy}},
  \bibinfo{journal}{Phys. Rev. B} \textbf{\bibinfo{volume}{84}},
  \bibinfo{pages}{035119} (\bibinfo{year}{2011}{\natexlab{b}}).

\bibitem[{\citenamefont{Nozi\`{e}res and Blandin}(1980)}]{nozieres}
\bibinfo{author}{\bibfnamefont{P.}~\bibnamefont{Nozi\`{e}res}}
  \bibnamefont{and} \bibinfo{author}{\bibfnamefont{A.}~\bibnamefont{Blandin}},
  \bibinfo{journal}{J. Phys. (Paris)} \textbf{\bibinfo{volume}{41}},
  \bibinfo{pages}{193} (\bibinfo{year}{1980}).

\bibitem[{\citenamefont{Logan et~al.}(2009)\citenamefont{Logan, Wright, and
  Galpin}}]{CW_multilevel}
\bibinfo{author}{\bibfnamefont{D.~E.} \bibnamefont{Logan}},
  \bibinfo{author}{\bibfnamefont{C.~J.} \bibnamefont{Wright}},
  \bibnamefont{and} \bibinfo{author}{\bibfnamefont{M.~R.}
  \bibnamefont{Galpin}}, \bibinfo{journal}{Phys. Rev. B}
  \textbf{\bibinfo{volume}{80}}, \bibinfo{pages}{125117}
  (\bibinfo{year}{2009}).

\bibitem[{\citenamefont{Wright et~al.}(2011)\citenamefont{Wright, Galpin, and
  Logan}}]{CW_multilevel2}
\bibinfo{author}{\bibfnamefont{C.~J.} \bibnamefont{Wright}},
  \bibinfo{author}{\bibfnamefont{M.~R.} \bibnamefont{Galpin}},
  \bibnamefont{and} \bibinfo{author}{\bibfnamefont{D.~E.} \bibnamefont{Logan}},
  \bibinfo{journal}{Phys. Rev. B} \textbf{\bibinfo{volume}{84}},
  \bibinfo{pages}{115308} (\bibinfo{year}{2011}).

\bibitem[{\citenamefont{Hofstetter and Schoeller}(2002)}]{HofSch}
\bibinfo{author}{\bibfnamefont{W.}~\bibnamefont{Hofstetter}} \bibnamefont{and}
  \bibinfo{author}{\bibfnamefont{H.}~\bibnamefont{Schoeller}},
  \bibinfo{journal}{Phys. Rev. Lett.} \textbf{\bibinfo{volume}{88}},
  \bibinfo{pages}{016803} (\bibinfo{year}{2002}).

\bibitem[{\citenamefont{Garst et~al.}(2004)\citenamefont{Garst, Kehrein,
  Pruschke, Rosch, and Vojta}}]{garst}
\bibinfo{author}{\bibfnamefont{M.}~\bibnamefont{Garst}},
  \bibinfo{author}{\bibfnamefont{S.}~\bibnamefont{Kehrein}},
  \bibinfo{author}{\bibfnamefont{T.}~\bibnamefont{Pruschke}},
  \bibinfo{author}{\bibfnamefont{A.}~\bibnamefont{Rosch}}, \bibnamefont{and}
  \bibinfo{author}{\bibfnamefont{M.}~\bibnamefont{Vojta}},
  \bibinfo{journal}{Phys. Rev. B} \textbf{\bibinfo{volume}{69}},
  \bibinfo{pages}{214413} (\bibinfo{year}{2004}).

\bibitem[{\citenamefont{Galpin et~al.}(2005)\citenamefont{Galpin, Logan, and
  Krishnamurthy}}]{CCDD2005}
\bibinfo{author}{\bibfnamefont{M.~R.} \bibnamefont{Galpin}},
  \bibinfo{author}{\bibfnamefont{D.~E.} \bibnamefont{Logan}}, \bibnamefont{and}
  \bibinfo{author}{\bibfnamefont{H.~R.} \bibnamefont{Krishnamurthy}},
  \bibinfo{journal}{Phys. Rev. Lett.} \textbf{\bibinfo{volume}{94}},
  \bibinfo{pages}{186406} (\bibinfo{year}{2005}).

\bibitem[{\citenamefont{Galpin et~al.}(2006)\citenamefont{Galpin, Logan, and
  Krishnamurthy}}]{CCDD2006}
\bibinfo{author}{\bibfnamefont{M.~R.} \bibnamefont{Galpin}},
  \bibinfo{author}{\bibfnamefont{D.~E.} \bibnamefont{Logan}}, \bibnamefont{and}
  \bibinfo{author}{\bibfnamefont{H.~R.} \bibnamefont{Krishnamurthy}},
  \bibinfo{journal}{J. Phys.: Condens. Matter} \textbf{\bibinfo{volume}{18}},
  \bibinfo{pages}{6545} (\bibinfo{year}{2006}).

\bibitem[{\citenamefont{Luttinger and Ward}(1960)}]{lutt_ward}
\bibinfo{author}{\bibfnamefont{J.~M.} \bibnamefont{Luttinger}}
  \bibnamefont{and} \bibinfo{author}{\bibfnamefont{J.~C.} \bibnamefont{Ward}},
  \bibinfo{journal}{Phys. Rev.} \textbf{\bibinfo{volume}{118}},
  \bibinfo{pages}{1417} (\bibinfo{year}{1960}).

\bibitem[{\citenamefont{Langreth}(1966)}]{lang}
\bibinfo{author}{\bibfnamefont{D.~C.} \bibnamefont{Langreth}},
  \bibinfo{journal}{Phys. Rev.} \textbf{\bibinfo{volume}{150}},
  \bibinfo{pages}{1516} (\bibinfo{year}{1966}).

\end{thebibliography}
\end{document}